\documentclass[sigconf]{acmart}

\def\BibTeX{{\rm B\kern-.05em{\sc i\kern-.025em b}\kern-.08emT\kern-.1667em\lower.7ex\hbox{E}\kern-.125emX}}
    
\usepackage{booktabs}
\usepackage{numprint}
\usepackage{acronym}
    
\usepackage{acronym}
\newacro{cots}[COTS]{Commercial Off The Shelf}   
\newacro{iiot}[IIoT]{Industrial Internet of Things}   
\newacro{iot}[IoT]{Internet of Things}     
\newacro{it}[IT]{Information Technology}
\newacro{ot}[OT]{Operation Technology}
\newacro{ip}[IP]{Internet Procotol}  
\newacro{hmi}[HMI]{Human Machine Interface}
\newacro{plc}[PLC]{Programmable Logic Controller}  
\newacro{cnc}[C\&C]{Command \& Control}  
\newacro{ics}[ICS]{Industrial Control System} 
\newacro{ids}[IDS]{Intrusion Detection System}
\newacro{scada}[SCADA]{Supervisory Control And Data Acquisition}  
\newacro{swat}[\textit{SWaT}]{\textit{Secure Water Treatment}}  
\newacro{lstm}[LSTM]{Long Short-Term Memory}
\newacro{cps}[CPS]{Cyber-Physical System}
\newacro{wsn}[WSN]{Wireless Sensor Networks}
\newacro{arima}[ARIMA]{Auto-Regressive Integrated Moving Average}
\newacro{dos}[DoS]{Denial of Service}    
\newacro{uv}[UV]{Ultraviolet}    
\newacro{ro}[RO]{Reverse Osmosis}    
\newacro{svm}[SVM]{Support Vector Machine}    
\newacro{sssp}[SSSP]{Single Stage Single Point}
\newacro{ssmp}[SSMP]{Single Stage Multi Point}
\newacro{mssp}[MSSP]{Multi Stage Single Point}
\newacro{msmp}[MSMP]{Multi Stage Multi Point}
\newacro{mtu}[MTU]{Master Terminal Unit}
\newacro{uf}[UF]{Ultra Filtration}
    
\copyrightyear{2019}
\acmYear{2019}
\setcopyright{acmlicensed}
\acmConference[ARES '19]{Proceedings of the 14th International Conference on 
Availability, Reliability and Security (ARES 2019)}{August 26--29, 
2019}{Canterbury, United Kingdom}
\acmBooktitle{Proceedings of the 14th International Conference on Availability, 
Reliability and Security (ARES 2019) (ARES '19), August 26--29, 2019, 
Canterbury, United Kingdom}
\acmPrice{15.00}
\acmDOI{10.1145/3339252.3341476}
\acmISBN{978-1-4503-7164-3/19/08}

\begin{document}

%
\title[Temporal and Topological Intrusion Detection in OT Networks]{Using Temporal and Topological Features for Intrusion Detection in Operational Networks}

%
\author{Simon D. Duque Anton}
\email{simon.duque_anton@dfki.de}
\orcid{0000-0003-4005-9165}
\affiliation{%
  \institution{German Research Center for AI}
  \streetaddress{Trippstadter Str. 122}
  \city{Kaiserslautern}
  \country{Germany}
  \postcode{67633}
}

\author{Daniel Fraunholz}
\email{daniel.fraunholz@dfki.de}
\affiliation{
  \institution{German Research Center for AI}
  \streetaddress{Trippstadter Str. 122}
  \city{Kaiserslautern}
  \country{Germany}
  \postcode{67633}
}

\author{Hans Dieter Schotten}
\email{hans_dieter.schotten@dfki.de}
\affiliation{
  \institution{German Research Center for AI}
  \streetaddress{Trippstadter Str. 122}
  \city{Kaiserslautern}
  \country{Germany}
  \postcode{67633}
}

%
\renewcommand{\shortauthors}{S. D. Duque Anton et al.}

\renewcommand\_{\textunderscore\allowbreak}

%
\begin{abstract}

Until two decades ago,
industrial networks were deemed secure due to physical separation from public networks.
An abundance of successful attacks proved that assumption wrong.
Intrusion detection solutions for industrial application need to meet certain requirements that differ from home- and office-environments,
such as working without feedback to the process and compatibility with legacy systems.
Industrial systems are commonly used for several decades,
updates are often difficult and expensive. 
Furthermore,
most industrial protocols do not have inherent authentication or encryption mechanisms,
allowing for easy lateral movement of an intruder once the perimeter is breached.
In this work,
an algorithm for motif discovery in time series,
\textit{Matrix Profiles},
is used to detect outliers in the timing behaviour of an industrial process.
This process was monitored in an experimental environment,
containing ground truth labels after attacks were performed.
Furthermore,
the graph representations of a different industrial data set that has been emulated are used to detect malicious activities.
These activities can be derived from anomalous communication patterns,
represented as edges in the graph.
Finally,
an integration concept for both methods is proposed.\\ \par
This is a preprint accepted at the 14th International Conference on Availability,
Reliability and Security (ARES 2019). Please cite as indicated in the ACM Reference Format below.
\end{abstract}

%
%


%

%
\maketitle

\section{Introduction}
Around the world,
enterprises are benefiting from the digitalisation.
The Industry 4.0 is promising reduced costs and time for acts like maintenance,
set up and configuration while increasing efficiency and thus revenue.
Similar to the \ac{iot} opening possibilities in end user environments,
the Industry 4.0 creates new business cases~\cite{Uckelmann.2011, Haller.2008}.
However,
the increase in interconnectivity and use of embedded intelligence on which the digitalisation of industry strives creates novel attack surfaces as well.
In the last decades,
attacks on industry have increased drastically~\cite{Duque_Anton.2017a}.
Opening network protocols that are not built for security to insecure public networks poses a threat to operation.
On the one hand,
criminals are targeting the cyber space,
and particularly the industrial sector.
Since 2007,
cyber crime creates a larger revenue than drug trafficking~\cite{Dethlefs.2015, Symantec.2009}.
On the other hand,
state-sponsored actors have allegedly taken a great interest in espionage and sabotage of critical infrastructure and industries,
with the Ukrainian power blackouts in December of 2015~\cite{Greenberg.2017} or \textit{Stuxnet}~\cite{Langner.2013} being only the widely-known examples.\\ \par
In order to protect industrial applications,
reliable intrusion detection methods and tools are necessary.
Researchers and vendors alike have taken an interest in industrial intrusion detection and prevention systems.
In contrast to home- and office applications,
where intrusion detection and prevention is well-established,
industrial applications pose different requirements on such solutions.
Legacy systems and protocols have to be integrated,
as well as often application specific entities in a network.
Downtimes are unacceptable as availability is the most important objective.
Updates are difficult due to the distributed nature of devices.
These requirements have to be considered by potential solutions.
In this work,
inherent features of industrial networks are made use of.
First,
behaviour during operation is predictable and regular.
Second,
the structure of a network is expected to remain stable for the most part.
Those characteristics are considered in a graph-based time series approach to detect attacks in a realistic data set created by an industrial use case.
First, 
a naive approach to time series is used to detect outliers.
After that, 
the traffic is represented in a graph-based way as to determine cause and effect of an attack.
In this application,
entity-specific time series-based analysis is performed.\\ \par
The contribution of this paper consists of three parts:
\begin{itemize}
\item The application of time series-based anomaly detection for detecting attacks in industrial process data is evaluated,
\item the application of graph-based anomaly detection of \ac{ip}-based indicators of intrusions in \ac{ot} networks is evaluated and
\item the concept of a combination of both approaches is discussed
\end{itemize}
The remainder of this work is structured as follows:
In Section~\ref{sec:sota},
related work is presented.
The data set and the industrial application it is derived from is discussed in Section~\ref{sec:data_set}.
After that,
the applied time series algorithm is introduced and evaluated in Section~\ref{sec:time_series}.
A structural analysis of an industrial attack scenario is discussed in Section~\ref{sec:graph_analysis}.
Both approaches are then combined in a concept for a holistic intrusion detection approach in Section~\ref{sec:hybrid_approach}.
Finally,
the findings are discussed in Section~\ref{sec:discussion}.

\section{Related Work}
\label{sec:sota}
In this section,
an overview of graph-based and time series-based anomaly detection in the context of intrusion detection is provided,
with a focus on industrial networks.
A summary of the related work is listed in Table~\ref{tab:sota}.
\begin{table}[h!]
\renewcommand{\arraystretch}{1.3}
\caption{Research Topics Covered by the Individual Works}
\label{tab:sota}
\centering
\scriptsize
\begin{tabular}{l  l}
\toprule
\textbf{Subject Covered} & \textbf{Scientific Work} \\
Reviews & \cite{Garcia-Teodoro.2008, Jyothsna.2011} \\
Graph-based methods & \cite{Akoglu.2014, Staniford-Chen.1996, Swiler.1998, Noble.2003, Eberle.2007, Pasqualetti.2013, Eswaran.2018} \\
Graph-based and time-sensitive methods \phantom{mm} & \cite{Tao.2018, Akoglu.2010b} \\
Machine learning-based & \cite{Debar.1992, Ma.2003, Ferdousi.2006} \\
Statistical processes & \cite{Moayedi.2008, Yaacob.2010, Tabatabaie.2016, Yu.2016} \\
Wavelet analysis & \cite{Munz.2007, Hamdi.2007, Lu.2009} \\
Industrial Intrusion Detection & \cite{Tsang.2005, Hadeli.2009, Fovino.2010, Regis_Barbosa.2010, Morris.2012, Gao.2014, Khalili.2015, Caselli.2015, Ponomarev.2016, Ghaeini.2016} \\
\bottomrule
\end{tabular}
\end{table}
There are reviews addressing the challenge of anomaly detection for intrusion detection.
\textit{Garc\'{i}a-Teodoro et al.} address the challenges of this field of work while presenting techniques and systems~\cite{Garcia-Teodoro.2008}.
\textit{Jyothsna and Prasad} provide a review of anomaly-based \acp{ids}~\cite{Jyothsna.2011}.
Graph-based methods for detecting anomalies contain different approaches.
\textit{Akoglu et al.} provide an exhaustive overview of the different kinds of methods~\cite{Akoglu.2014}.
They distinguish between static and dynamic graphs. 
In static graphs,
the goal is to identify nodes or edges that are significantly different from the rest of nodes or edges respectively.
In dynamic graphs,
the goal is to compare the object under observation in a graph representation in different time steps.
If at any time the characteristics differ,
an outlier shall be detected.
In 1996,
\textit{Staniford-Chen et al.} present an \ac{ids} for large networks,
based on a graph representation of said network~\cite{Staniford-Chen.1996}.
The network information is aggregated into activity graphs,
making it feasible to detect coordinated attacks.
This system is presented in more detail in a succeeding work~\cite{Cheung.1999}.
\textit{Swiler and Phillips} present a graph-based system for network vulnerability analysis~\cite{Swiler.1998}.
The general idea lies in the assignment of properties to each node and edge in the network under observation.
This is mapped to possible attack vectors in the network,
resulting in a probability for each type of attack on each asset in the network.
\textit{Noble and Cook} introduce novel methods for anomaly detection in graphs~\cite{Noble.2003}.
Additionally,
they present a metric for the regularity of a graph,
indicating the probability of an outlier.
This method is extended by \textit{Eberle and Holder}~\cite{Eberle.2007}.
They add types of anomalies to the model and provide methods for detecting different kinds of anomalies at different places in the graph.
\textit{Pasqualetti et al.} present a method specifically for detecting attacks on \acp{cps} while considering graph-properties~\cite{Pasqualetti.2013}.
A different kind of approach for detecting anomalies in graph is presented by \textit{Eswaran and Faloutsos}~\cite{Eswaran.2018}.
They consider dynamic graphs and look at the edges in any given time step,
called an edge stream.
This stream of edges is then used to detect anomalies by identifying bogus edges.
\textit{Tao et al} consider graph representations of online accounts~\cite{Tao.2018}.
They extract patterns,
use them to train deep neural networks such as \acp{lstm} and employ these to detect fraudulent takeover of accounts.\\ \par
Time series-based anomaly detection is used to detect outliers in a data series that is sequential in time.
These outliers are usually anomalous in comparison to the previous values of the time series.
After creating time series from a data set,
there are different ways to analyse this data.
\textit{Akoglu and Faloutsos} combine graph-based and time series-based anomaly detection~\cite{Akoglu.2010b}
They consider mobile communication networks and analyse the behaviour of communication over time.
If entities in the network change their behaviour,
they are detected as responsible for anomalies.
One way of analysing time series is by using it to train neural networks.
\textit{Debar et al.} present a neural network-based approach in 1992~\cite{Debar.1992}.
Even though they do not explicitly mention the concept of time series,
they use a neural network to learn normal behaviour in a network and flag deviations as attacks.
\textit{Ma and Perkins} employ one-class \acp{svm} to find anomalies in time series~\cite{Ma.2003}.
\textit{Ferdousi and Maeda} employ unsupervised outlier detection techniques on time series data,
namely peer group analysis~\cite{Ferdousi.2006}.\\ \par
Other than that,
statistical processes are used to model the timing behaviour.
Consequently,
an anomaly is detected if the observed behaviour does not match the modeled values.
A common approach is using \ac{arima} to model time series behaviour~\cite{Moayedi.2008, Yaacob.2010},
\textit{Tabatabaie et al.} include a chaotic behaviour prediction into their \ac{arima} model~\cite{Tabatabaie.2016}.
\textit{Yu et al.} present an anomaly detection scheme based on \ac{arima} for \acp{wsn}~\cite{Yu.2016}.\\ \par
Apart from \ac{arima} models,
wavelet analysis has been employed for detecting anomalies in network traffic~\cite{Lu.2009},
in flows~\cite{Munz.2007}.,
i.e. aggregated network traffic information and to detect \ac{dos} attacks~\cite{Hamdi.2007}.\\ \par
Intrusion detection in the industrial domain is specific with respect to certain parameters.
Legacy systems without inherent security mechanisms have to be addressed~\cite{Fovino.2010, Morris.2012, Gao.2014},
critical states that can have severe effects on the physical world need to be prevented~\cite{Khalili.2015} and deterministic behaviour of processes can be leveraged to detect anomalies~\cite{Hadeli.2009}.
Sequences are relatively unifom in industrial applications,
this characteristic can be incorporated into an \ac{ids}~\cite{Caselli.2015}.
\textit{Tsang and Kwong} present an industrial \ac{ids} based on the ant colony clustering approach~\cite{Tsang.2005}.
\textit{Regis Barbosa and Pras} present a novel flow-based intrusion and anomaly detection method~\cite{Regis_Barbosa.2010}.
Air gapped \acp{ics} and attacks on such systems are evaluated by \textit{Ponomarev and Atkison}~\cite{Ponomarev.2016}.
\textit{Ghaeini and Tippenhauer} present a hierarchical model for industrial intrusion detection to combine information from the physical,
as well as the \ac{plc} layer~\cite{Ghaeini.2016}.

\section{Data Set}
\label{sec:data_set}
\begin{figure}
  \includegraphics[width=\linewidth]{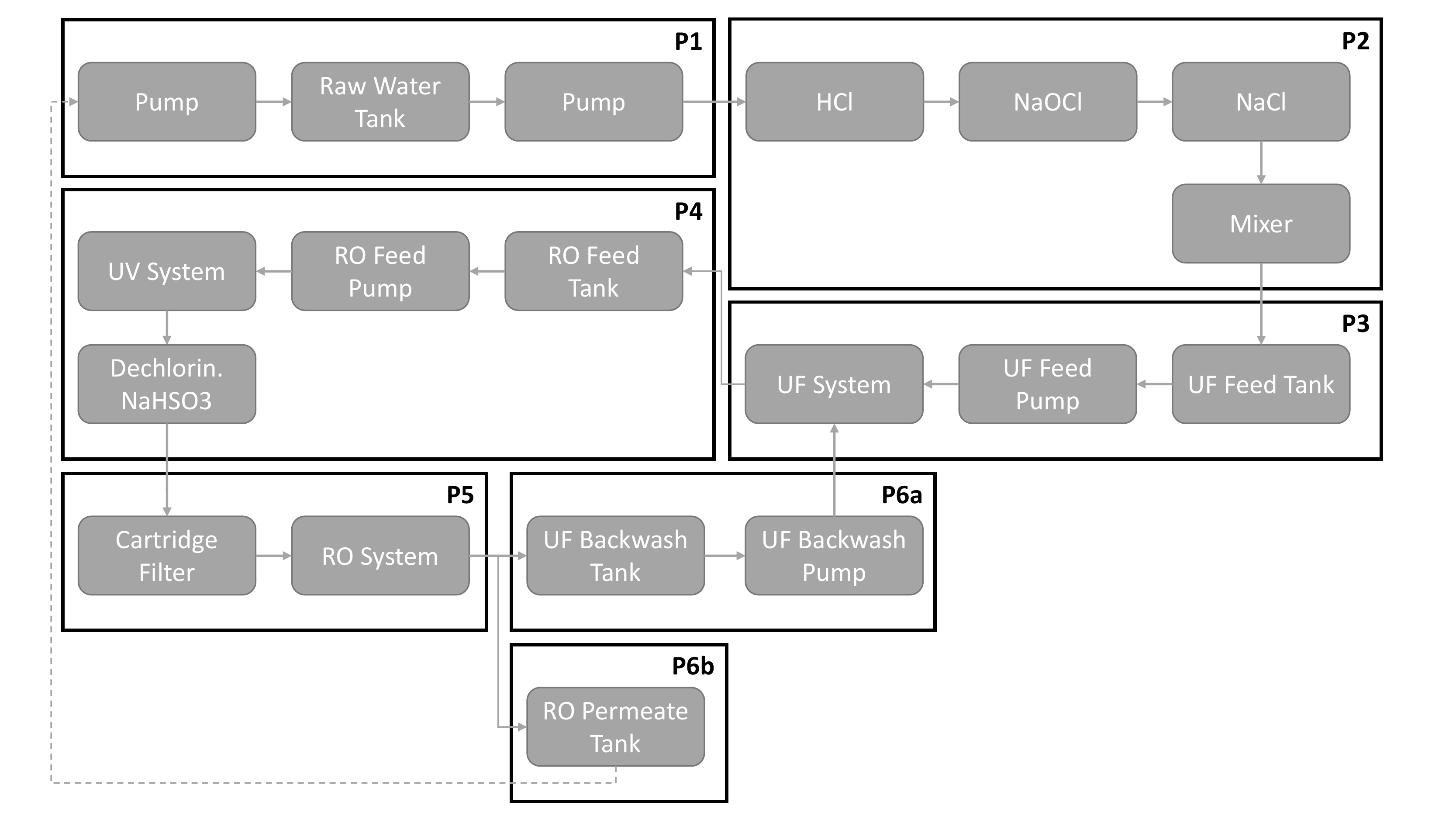}
  \caption{Relation of Sub-Processes}
  \label{fig:process_order}
\end{figure}

In this work,
a data set containing network and process data of a research facility is investigated.
The process has been monitored for a total of eleven days,
where seven days were ran as normal operation,
while the four last days contained attacks.
The data set used in this work is provided by \textit{iTrust,
Centre for Research in Cyber Security,
Singapore University of Technology and Design} and is titled \ac{swat}~\cite{Goh.2016, iTrust.2018}.
It consists of pcap-files containing the packets of the \ac{ot} network traffic as well as csv-lists containing the sensor and actuator values at each time point.
It has been widely used in scientific reserach,
e.g. by \textit{Schneider and B\"ottinger}~\cite{Schneider.2018}.
The process contains six different sub-processes, 
controlled by one \ac{plc} each.
In the course of the process,
raw water is stored, 
assessed for its quality and treated by different methods.
The sub-processes are:
\begin{itemize}
\item \textit{P1}: Raw water storage
\item \textit{P2}: Pre-treatment
\item \textit{P3}: Membrane \ac{uf}
\item \textit{P4}: Dechlorination by \ac{uv} lamps
\item \textit{P5}: \ac{ro}
\item \textit{P6}: Disposal
\end{itemize}
These sub-processes are connected as described in Figure~\ref{fig:process_order}.
First,
the raw water is stored in a tank.
It is treated by initial measures.
After that,
it is filtered and treated with \ac{uv} light and \ac{ro}.
If it is sufficiently clean,
it is stored in the final tank.
If not, 
the \ac{uf} and \ac{uv} treatment are repeated. \\ \par
\begin{figure}
  \includegraphics[width=\linewidth]{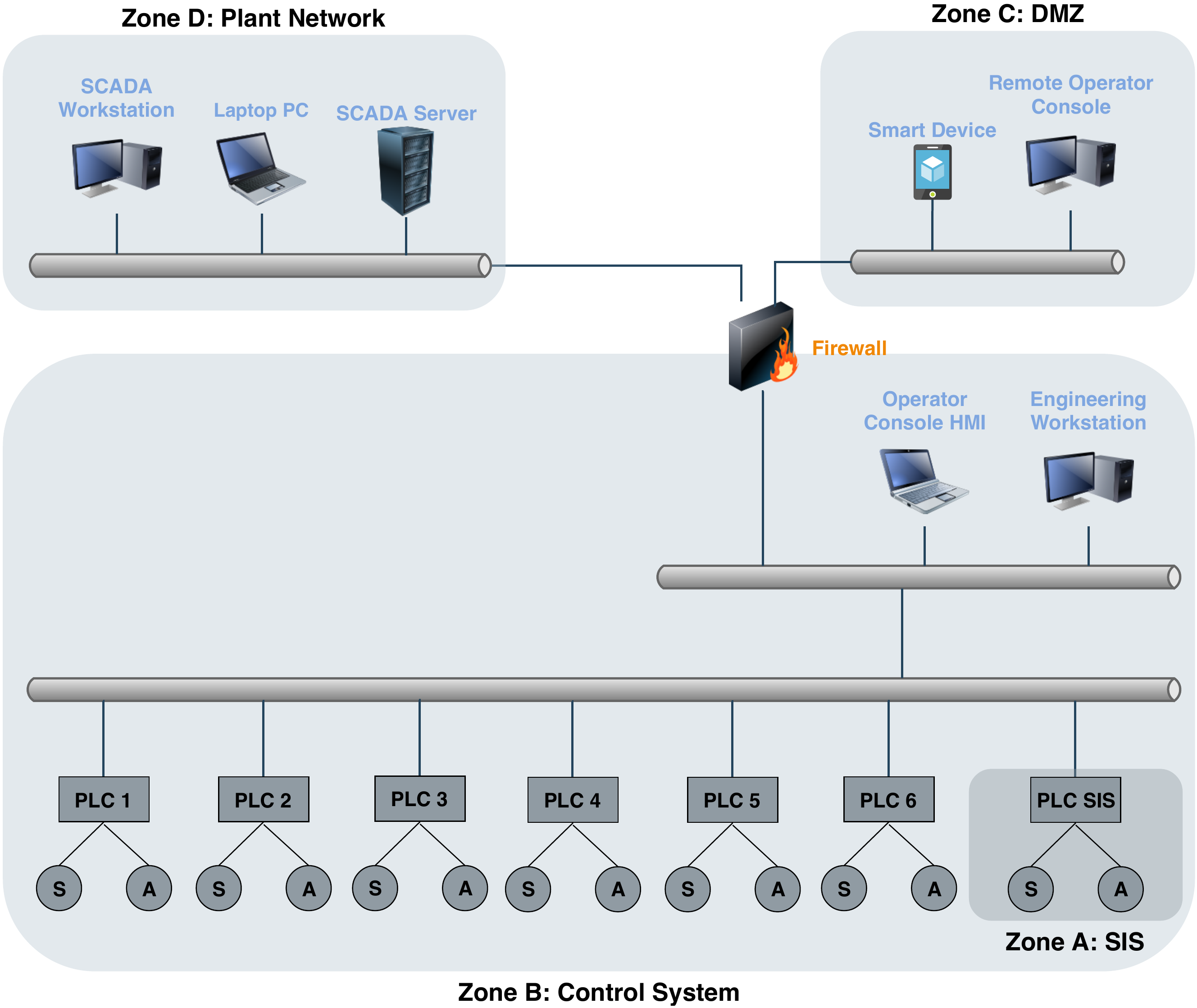}
  \caption{Schematic Overview of the Process Environment}
  \label{fig:nw_structure}
\end{figure}

Each sub-process is controlled by one \ac{plc}.
These \acp{plc} control sensors and actuators in a ring network.
They are in turn controlled and monitored by \acp{hmi},
a \ac{scada} workstation as well as a data historian in a star network.
A schematic representation of the communication relations can be found in Figure~\ref{fig:nw_structure}.
After seven days of normal operation,
a total of 41 attacks was introduced to the process.
All data is labeled,
providing ground truth for the data.
The operators distinguish four different kinds of attacks~\cite{Goh.2016}:
\begin{itemize}
\item \textit{\ac{sssp}}: Single stage attack on one point in the process, 26 instances in the data set
\item \textit{\ac{ssmp}}: Single stage attack on multiple points in the process, 4  instances in the data set
\item \textit{\ac{mssp}}: Multi stage attack on one point in the process, 2 instances in the data set
\item \textit{\ac{msmp}}: Multi stage attack on multiple points in the process, 4 instances in the data set
\end{itemize}
Of the total 41 attacks,
18 did not create an actual change in one of the sub-processes. \\ \par
In total,
51 sensors and actuators were controlled by the six \acp{plc}.
The exhaustive list and functionality can be found in the work of \textit{Goh et al.}~\cite{Goh.2016}.
The attacks were executed on 27 of these sensor and actuators.
In Table~\ref{tab:source},
the top five of affected sensors are listed,
their function is described as well as the number of attacks executed on it, 
including the number of attacks that did not result in a change in the process.
\begin{table}[h!]
\renewcommand{\arraystretch}{1.3}
\caption{Sources of the Attacks}
\label{tab:source}
\centering
\scriptsize
\begin{tabular}{l l l r r}
\toprule
\textbf{Elem.} & \textbf{Sub-Process} & \textbf{Description} & \textbf{Total} & \textbf{No Change} \\
P-102 & P1 & Pump (backup) \phantom{mi} & 3 & \phantom{mmmmmmm} 0 \\
P-101 & P1 & Pump & 2 & 0 \\
MV-101 \phantom{mi} & P1 \phantom{mmmmmmm} & Motor valve  & 2 & 0 \\
P-203 & P2 & Dosing pump & 2 & 0 \\
P-302 & P3 & \ac{uf} feed pump & 2 & 0 \\
\bottomrule
\end{tabular}
\end{table}
Apart from the source,
each attack could be detected at one point in the system. 
The top five points of detection,
their descriptions,
as well as the numbers of attacks that should affect them and the numbers that actually did,
are shown in Table~\ref{tab:detectable}.
\begin{table}[h!]
\renewcommand{\arraystretch}{1.3}
\caption{Detectable Points of Attacks}
\label{tab:detectable}
\centering
\scriptsize
\begin{tabular}{l  l  l r r}
\toprule
\textbf{Elem.} & \textbf{Sub-Process} & \textbf{Description} & \textbf{Total} & \textbf{No Change} \\
LIT-101 & P1 & Raw water tank level \phantom{mi} & 7 & \phantom{mmmmmmm} 3 \\
P-101 & P1 & Pump & 2 & 0 \\
LIT-301 \phantom{mi} & P3 \phantom{mmmmmmm} & \ac{uf} feed tank level & 5 & 3 \\
MV-303 & P3 & Motorised valve & 2 & 0 \\
LIT-401 & P4 & \ac{ro} feed tank level & 3 & 1 \\
\bottomrule
\end{tabular}
\end{table}

\section{Time Series Analysis: \textit{Matrix Profiles}}
\label{sec:time_series}
In this section, 
a relatively novel method for time series analysis is applied to the data set described in Section~\ref{sec:data_set}:
The \textit{Matrix Profiles} approach~\cite{Yeh.2016a}.
It was introduced by \textit{Yeh et al.} in 2016 and has received many extensions since.\\ \par
The concept of \textit{Matrix Profiles} is to calculate the distance of any part of the time series,
called \textit{motif},
from any other motif in the time series.
Then, 
the minimal distance of the given motif from any other one is used as the \textit{Matrix Profile}.
A low \textit{Matrix Profile} indicates at least one similar motif in the time series,
while a high \textit{Matrix Profile} indicates an outlier.
In the context of this work,
outliers are of interest.
A general assumption about industrial processes is a regularity in the timely behaviour.
Consequently,
any attack that disrupts this regular behaviour will be detected as it creates motifs that are unique.
One of the non-trivial challenges of anomaly-based intrusion detection is distinguishing between non-malicious and malicious attacks.
However,
for the course of this work and as a generalisation,
any disturbance in process behaviour is worth investigating. 
Only the consequence should differ,
i.e. the detection of an attack should result in defense mechanisms,
while non-malicious anomalies should result in maintenance efforts.
Furthermore,
there can be slow changes in process behaviour that are normal,
as well as abrupt changes due to reconfiguration of the process.
The former can be addressed by \textit{Matrix Profiles}~\cite{Zhu.2017},
while the latter can be addressed by annotated \textit{Matrix Profiles}~\cite{Dau.2017},
as reconfigurations are expected to be known beforehand.\\ \par
\textit{Matrix Profiles} require one hyperparameter,
$m$,
the length of the motifs.
This hyperparameter,
however,
is robust to changes.
We found that the highest occurring frequency in a time series signal is a good guess.
In the context of this work,
two thirds of a day of normal behaviour was used as a baseline for the \textit{Matrix Profiles}.
Any possible deviations and irregularities in the normal behaviour are expected to occur during that time.
Furthermore,
about three days of process during which attacks occurred were analysed,
namely the 29.12.2015 at 18:09:28 until the 31.12.2015 at 02:36:40.
During this time,
ten attacks,
which are supposed to be detectable at twelve sensors and actuators,
occur.
The reason for not taking into account all of the available data lies in the size:
In the course of this work,
more than \numprint{230000} time points were analysed.
Each of the time points contains the values of 51 sensors and actuators.
In order to evaluate the data set,
sensors and actuators that are affected by the given attacks are identified and evaluated.
However,
\textit{Matrix Profiles} do have an extension that provides multidimensionality.
We used the single-dimensional approach for better visualisation and interpretation of the data,
an automated approach would consider many dimensions at once. \\ \par
We considered 13 sensors and actuators to evaluate for indicators of attacks:
\begin{itemize}
\item AIT-502: Sensor, measures NaOCl-level in \ac{ro}-subprocess
\item DPIT-301: Sensor, measures differential pressure in backwash-subprocess
\item LIT-101: Sensor, measures raw water tank level
\item LIT-301: Sensor, measures \ac{uf} water tank level
\item LIT-401: Sensor, measures \ac{uf} water tank level
\item MV-101: Actuator, controls water flow to raw water tank
\item MV-201: Actuator, controls water flow to \ac{uf} water tank
\item P-101: Actuator, pumps water from raw water tank to second subprocess
\item P-203: Actuator, HCl dosing pump
\item P-205: Actuator, NaOCl dosing pump
\item P-302: Actuator, pumps water from \ac{uf} subprocess to \ac{ro} suprocess
\item P-501: Actuator, pumps dechlorinated water to \ac{ro}
\item UV-401: Actuator, removes chlorine from water
\end{itemize}
However,
only AIT-502 to LIT-401 provided sensible results.
The other sensors and actuators are binary in their value range.
Unfortunately,
\textit{Matrix Profiles} do not work well with digital values.
The z-normalisation is not meant for motifs with zero standard deviation which occur frequently in case of 0 and 1.
Extensions are possible for boolean values encoded as binary digital,
creating motifs and calculating distances based on different distance metrics.
However,
they are not in the scope of this work and are left as future extensions. \\ \par
\begin{figure}[!ht]
\centering
\includegraphics[width=\linewidth]{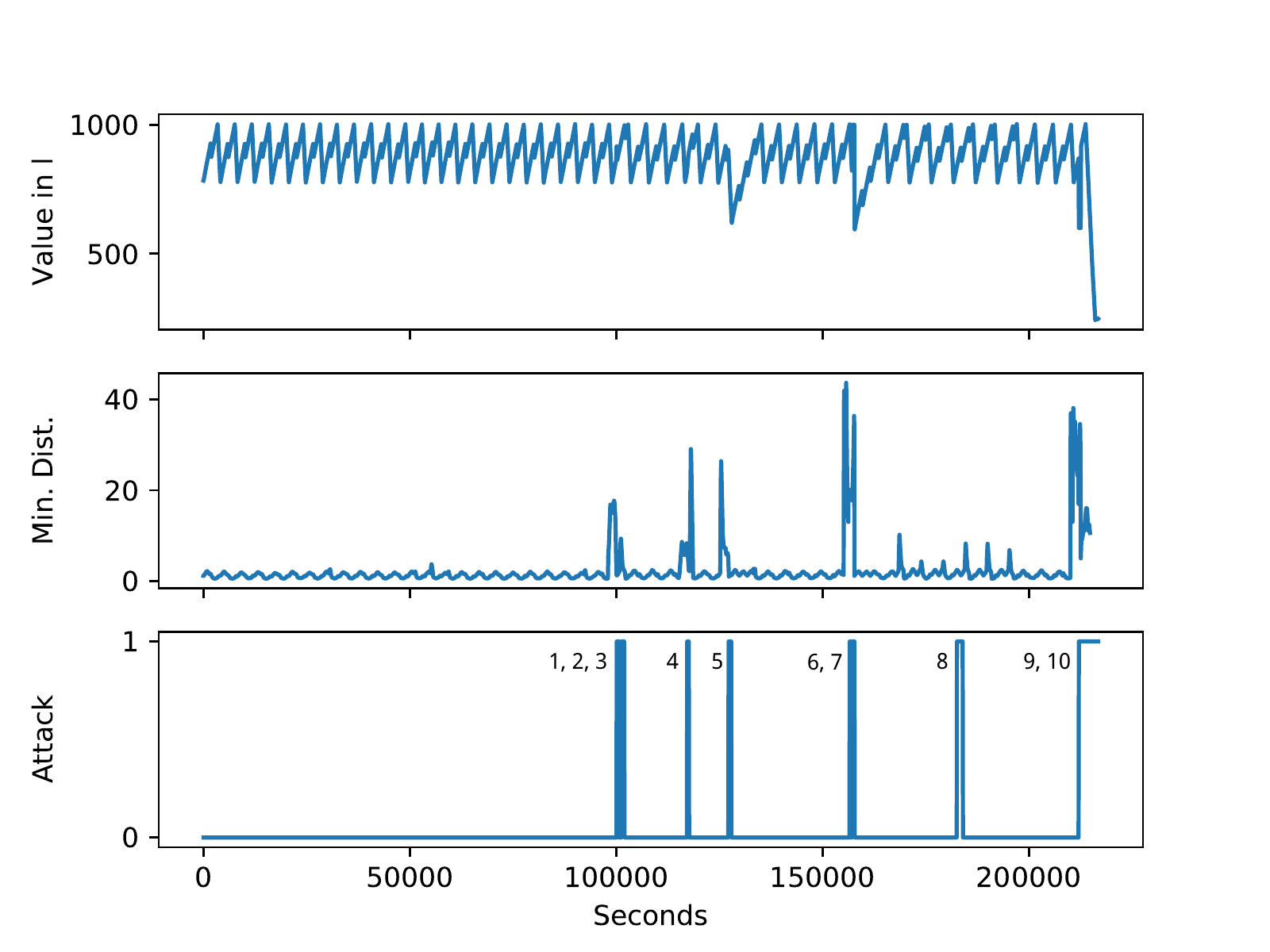}
\caption{\textit{Matrix Profile} of LIT-401}
\label{fig:LIT-401}
\end{figure}
\begin{figure}[!ht]
\centering
\includegraphics[width=\linewidth]{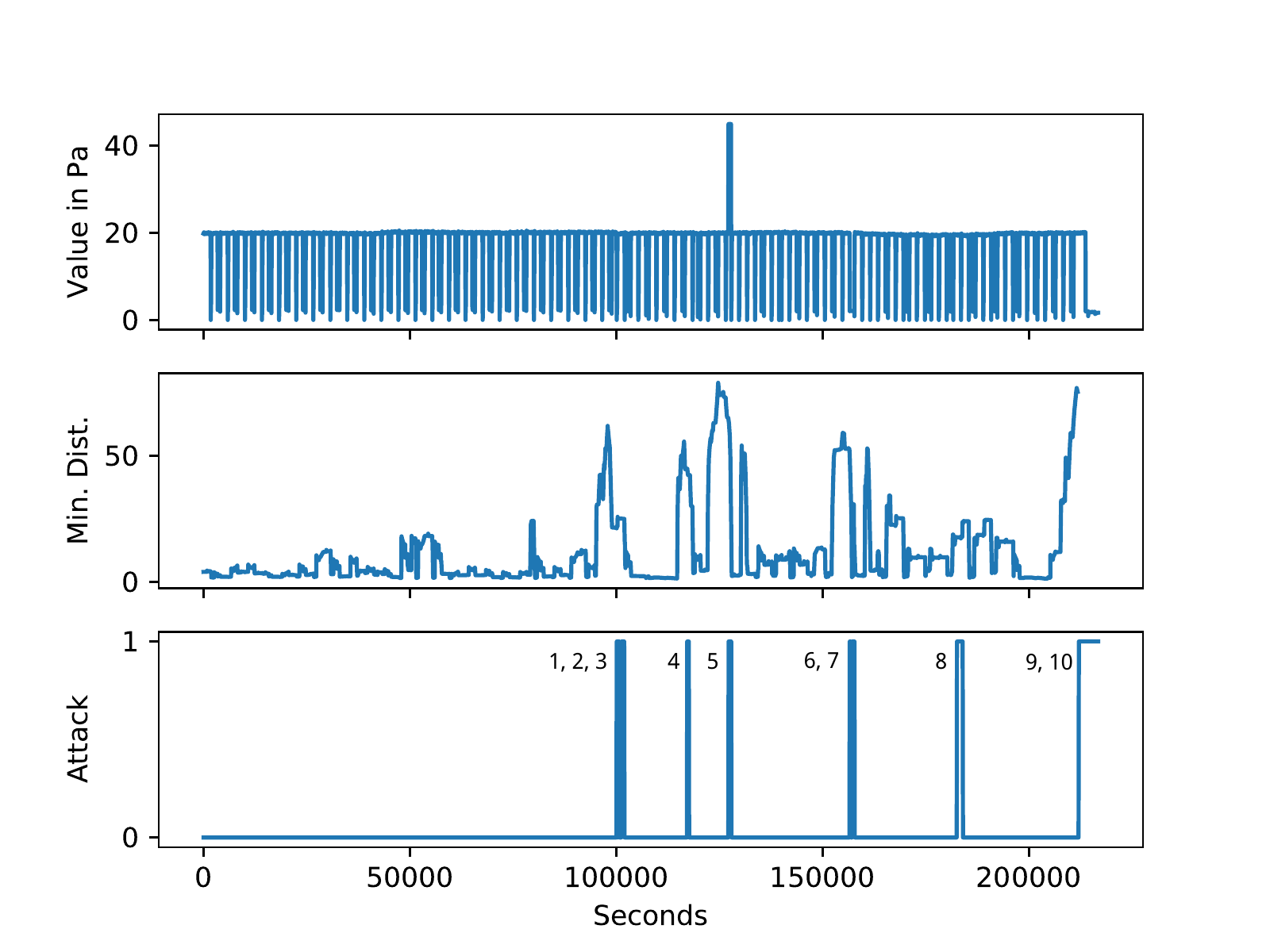}
\caption{\textit{Matrix Profile} of DPIT-301}
\label{fig:DPIT-301}
\end{figure}
\begin{figure}[!ht]
\centering
\includegraphics[width=\linewidth]{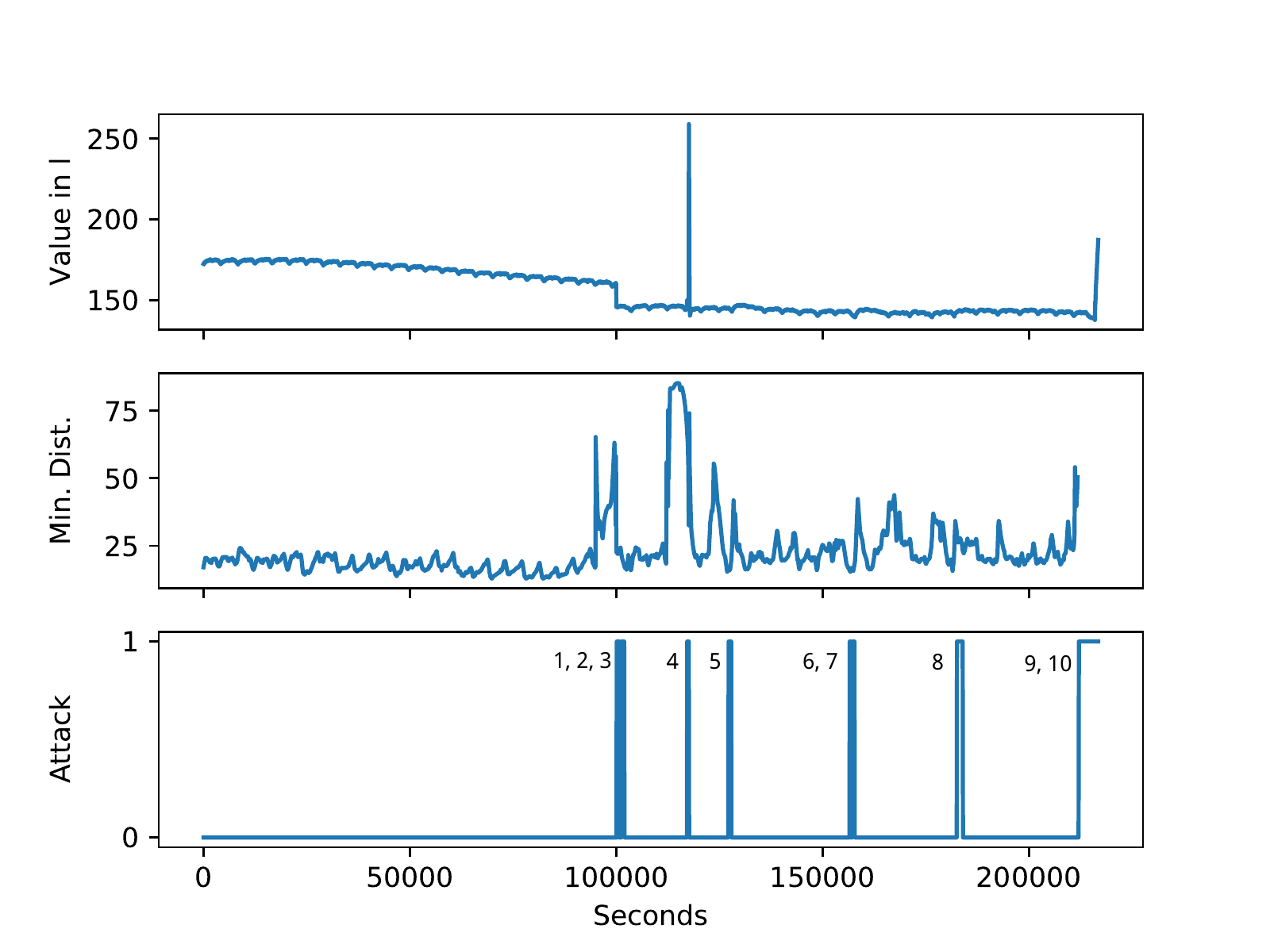}
\caption{\textit{Matrix Profile} of AIT-502}
\label{fig:AIT-502}
\end{figure}

The attacks occurring during the evaluated time period are listed in Table~\ref{tab:attack_during_eval},
including the source and result of the attack.
\begin{table}[h!]
\renewcommand{\arraystretch}{1.3}
\caption{Attacks During Evaluation}
\label{tab:attack_during_eval}
\centering
\scriptsize
\begin{tabular}{r l l}
\toprule
\textbf{Attack number} & \phantom{mi} \textbf{Source} & \textbf{Detectable} \\
1 & \phantom{mi} AIT-504 & AIT-504\\
2 & \phantom{mi} AIT-504 & AIT-504\\
3 & \phantom{mi} MV-101, LIT-101 & LIT-101 \\
4 & \phantom{mi} UV-401, AIT-502, P-501 & UV-401, AIT-502, P-501 \\
5 & \phantom{mi} P-602, DIT-301, MV-302 \phantom{mi} & DPIT-301, FIT-301 \\
6 & \phantom{mi} P-203, P-205 & P-203, P-205 \\
7 & \phantom{mi} LIT-401, P-401 & LIT-401 \\
8 & \phantom{mi} P-101, LIT-101 & LIT-301, LIT-101 \\
9 & \phantom{mi} P-302, LIT-401 & LIT-401 \\
10 & \phantom{mi} P-302 & LIT-401, P-302 \\
\bottomrule
\end{tabular}
\end{table}
It should be noted that attacks number 1, 2 and 3,
attacks number 6 and 7,
and attacks number 9 and 10 respectively are indistinguishable in the following figures due to their appearances shortly after one another.
In the following figures,
the attack numbers are noted next to the corresponding events.
Even though the attacks are supposed to be detectable only at certain points in the system,
LIT-401 can be used to successfully detect seven attacks,
shown in Figure~\ref{fig:LIT-401}.
The attacks are annotated in the lower signal,
one indicating an attack according to the label.
This constitutes the visualisation of the ground truth for evaluation purposes.
The middle line shows the \textit{Matrix Profile},
i.e. the minimal distance of the motif at this position with length $m$ from any other motif.
For this sensor,
$m$ was set to 500 seconds.
It can be seen that seven of the attacks correspond to drastic increases in the minimal distance,
indicating anomalous behaviour.
The sensor LIT-301 can be used to detect attack number eigth,
indicated by the second right peak in the attack-line of Figure~\ref{fig:LIT-301}.
\begin{figure}[!ht]
\centering
\includegraphics[width=\linewidth]{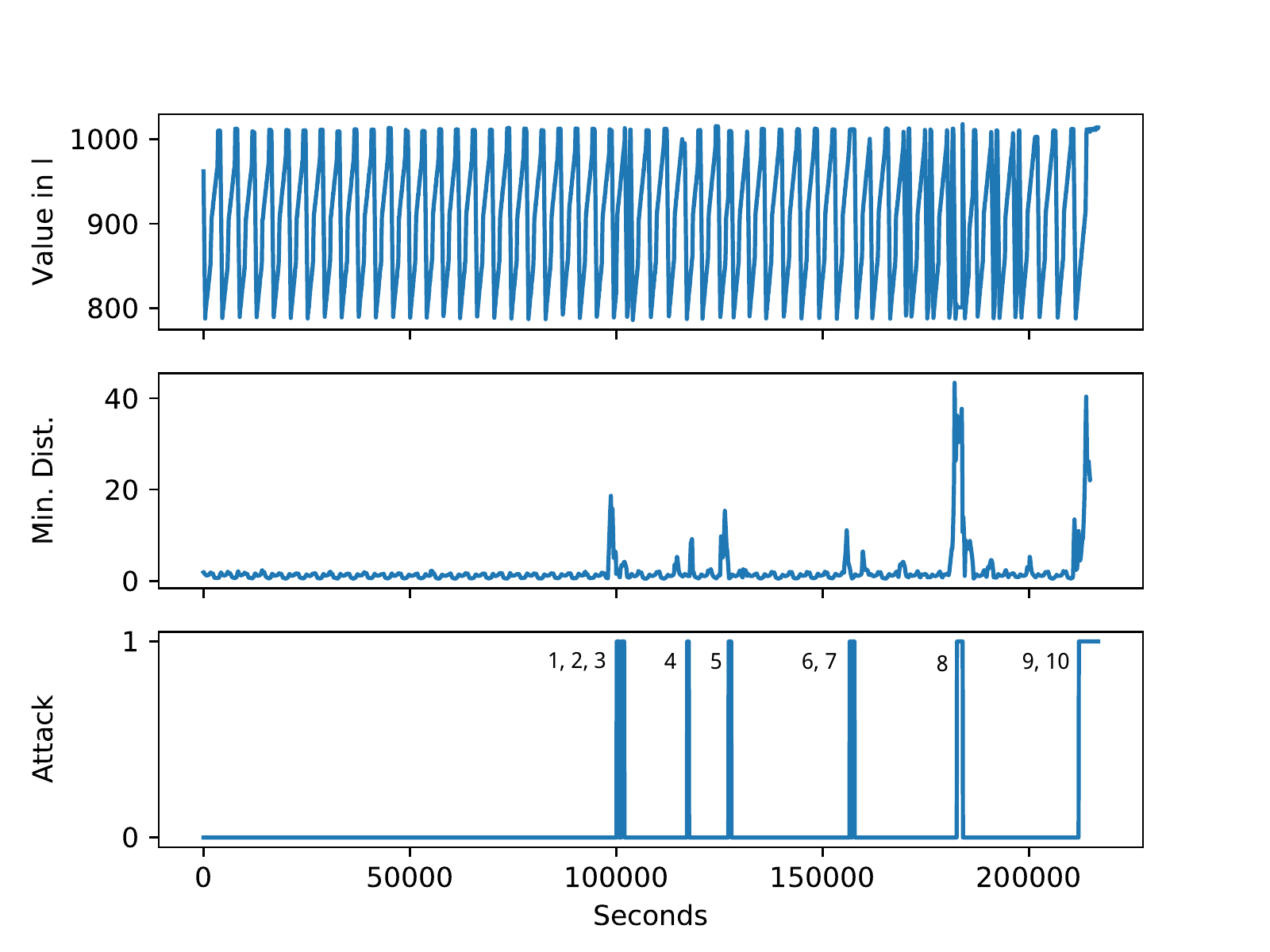}
\caption{\textit{Matrix Profile} of LIT-301}
\label{fig:LIT-301}
\end{figure}
The sensor LIT-101 can be used to detect every attack except for attack number 4,
indicated by the second left peak in the lower line.
This is shown in Figure~\ref{fig:LIT-301}.
\begin{figure}[!ht]
\centering
\includegraphics[width=\linewidth]{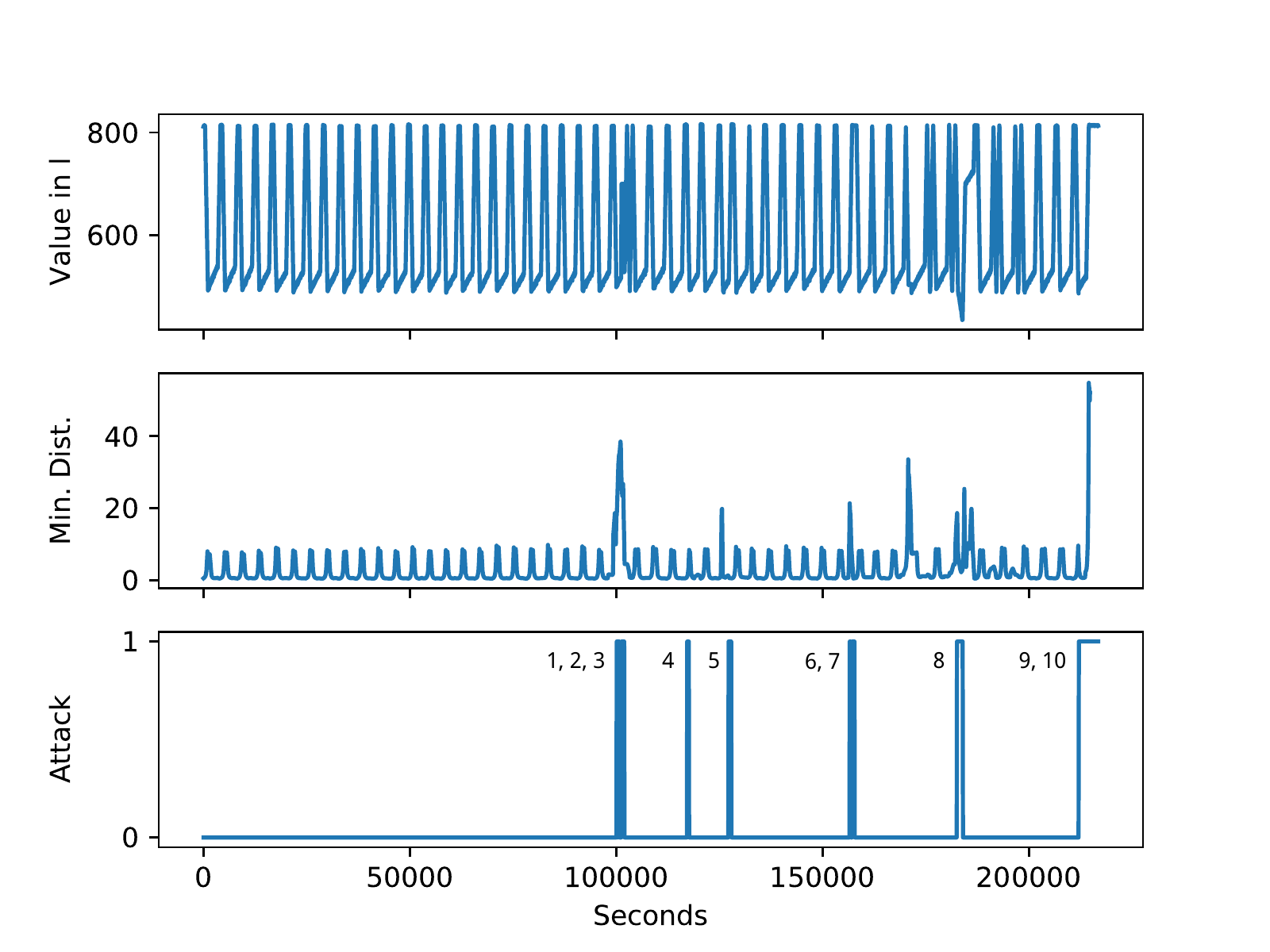}
\caption{\textit{Matrix Profile} of LIT-101}
\label{fig:LIT-301}
\end{figure}
The sensors DPIT-301 and AIT-502 can be used to detect attacks as well,
shown in Figures~\ref{fig:DPIT-301} and~\ref{fig:AIT-502}.
\begin{figure}[!ht]
\centering
\includegraphics[width=\linewidth]{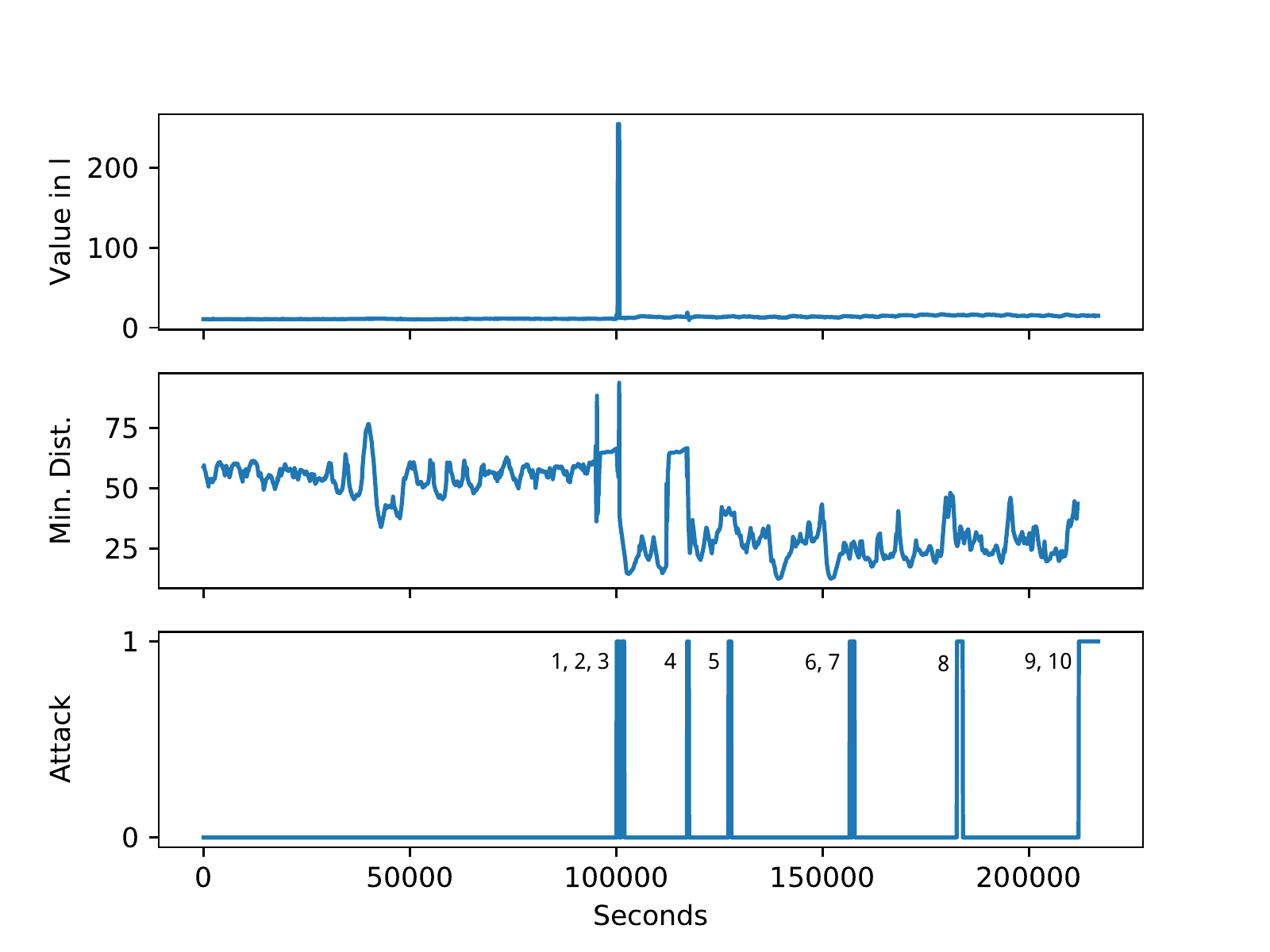}
\caption{\textit{Matrix Profile} of AIT-504}
\label{fig:AIT-504}
\end{figure}
However,
$m$ was increased in these cases.
Even though it is robust to change,
large differences between period and $m$ lead to suboptimal performance.
The autocorrelation analysis of DPIT-301 shows a period at 2000 seconds,
thus,
$m$ was set to 2000 seconds.
As a rule of thumb,
$m$ has proven to need to at least exceed the first period of the time series data under investigation.
This knowledge can be used by automated anomaly detection algorithms as well,
since it is easy to compute.
Starting from this value,
the quality of attack detection increases.
Preliminary analyses indicate that a value of $m$ of the first period or larger significantly improves the performance of \textit{Matrix Profiles}.
AIT-502 does not have a clear period,
but the quality of detection also increases with $m$.
In this evaluation,
a value of 5000 seconds was chosen for $m$.
AIT-504 shows constant behaviour with small deviations and one definite outlier,
depicted in Figure~\ref{fig:AIT-504}.
The comparably high values for the minimal distances during the training period make detection of the attacks difficult.
According to the description of the data set~\cite{iTrust.2018},
the first two attacks affect this sensor.
Despite the variations,
this constitutes the highest minimal distance.

\section{Graph-based Analysis}
\label{sec:graph_analysis}
The data set analysed in the previous sections contains attacks that are based on tampered process parameter.
Sensor and actuator values are changed in order to disrupt the process flow.
However,
the attack vector that was used to gather access to said sensors and actuators is not described.
Assumptions could be made,
about attackers breaching air-gapped \ac{scada}-networks of industrial applications.
Since no traces of breaches can be found in the monitored network traffic,
these assumptions do not aid intrusion detection mechanisms. \\ \par
Due to the strict topology of industrial networks,
communication patterns can be employed to detect attacks as well.
\textit{Lemay and Fernandez} created a set of network traffic they monitored in an emulated environment~\cite{Lemay.2016}.
This set of network traffic contains of pcap-files containing the \ac{ot} network packets for different set-ups.
After simulating an industrial process consisting of circuit breakers,
they introduce different kinds of attacks into the system.
These attacks do not employ specifics of the used \textit{Modbus} protocol,
but instead are TCP/IP-based attacks conducted with \textit{metasploit}.
That means process-based features are of no use to detect attacks.
However,
network packet characteristics~\cite{Duque_Anton.2018b},
as well as meta-data information as a time-series~\cite{Duque_Anton.2018c} can be used to successfully detect attacks. 
The number of port- and IP-pairs has a strong impact on the detection in doing so.\\ \par
\begin{figure}[!ht]
\centering
\includegraphics[width=\linewidth]{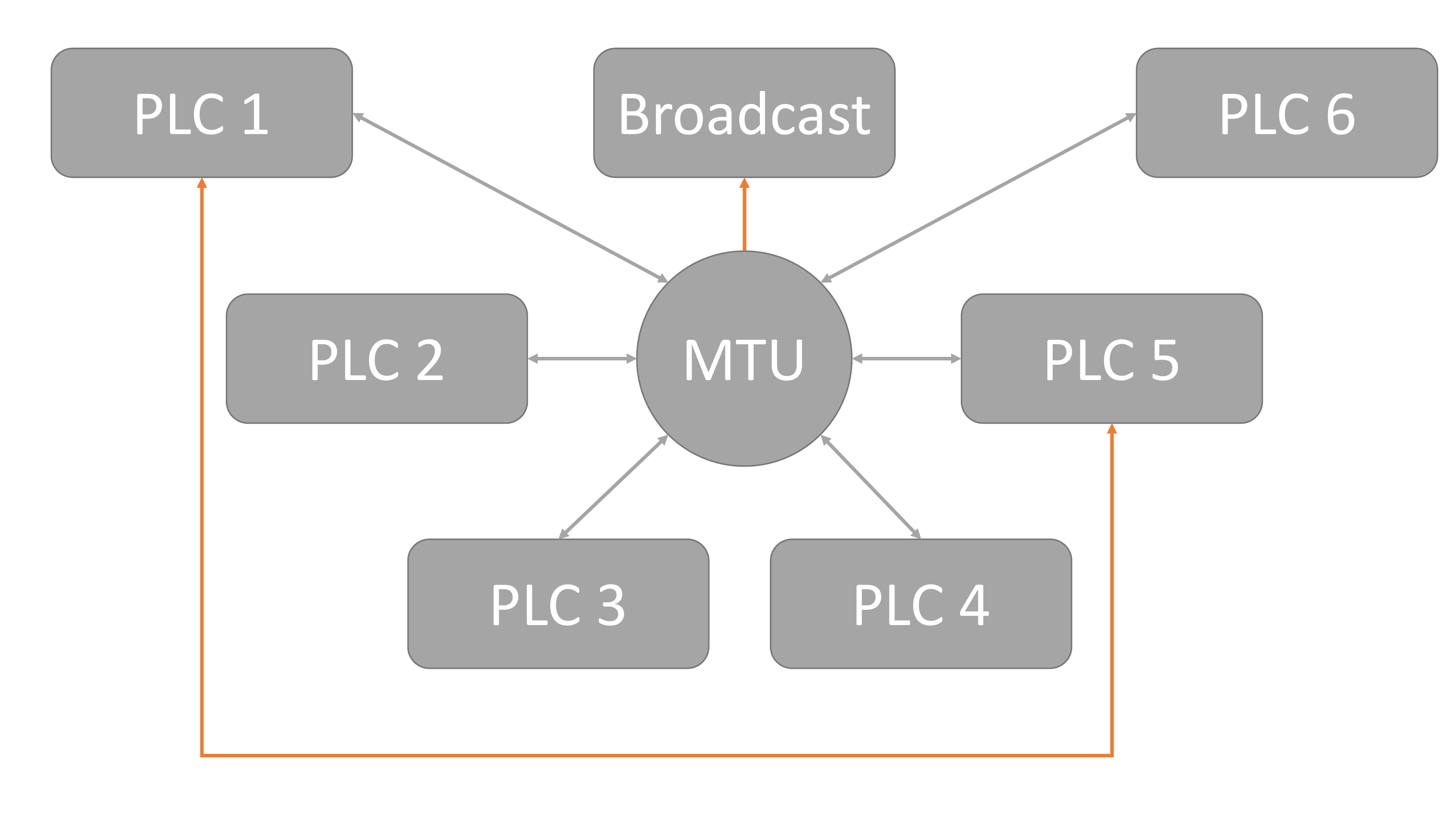}
\caption{Communication Structure of Data Set \textit{CnC\_uploading\_exe\_modbus\_6RTU\_with\_operate}}
\label{fig:graph1}
\end{figure}
\begin{figure}[!ht]
\centering
\includegraphics[width=\linewidth]{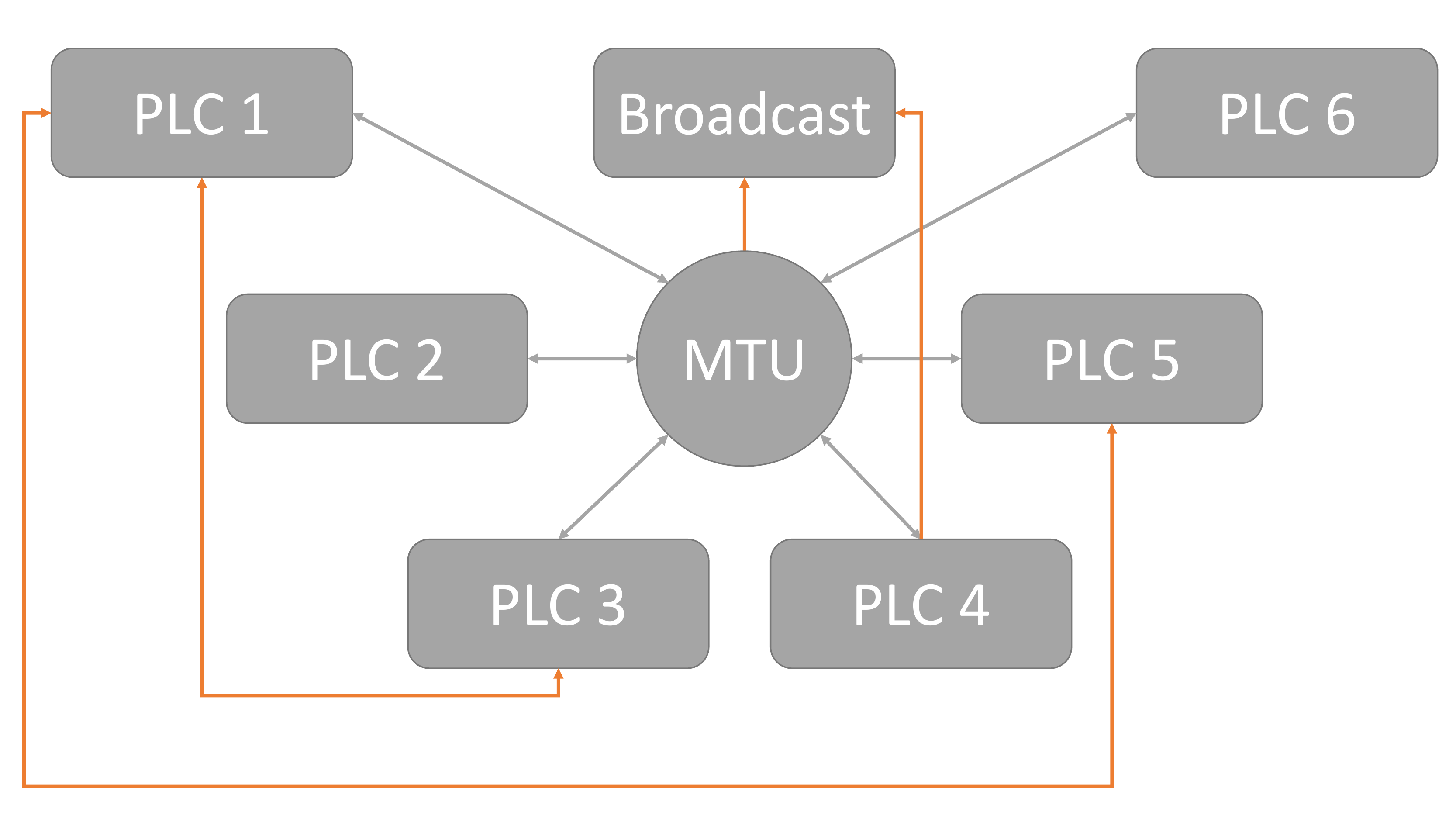}
\caption{Communication Structure of Data Set \textit{Moving\_two\_files\_Modbus\_6RTU}}
\label{fig:graph2}
\end{figure}
\begin{figure}[!ht]
\centering
\includegraphics[width=\linewidth]{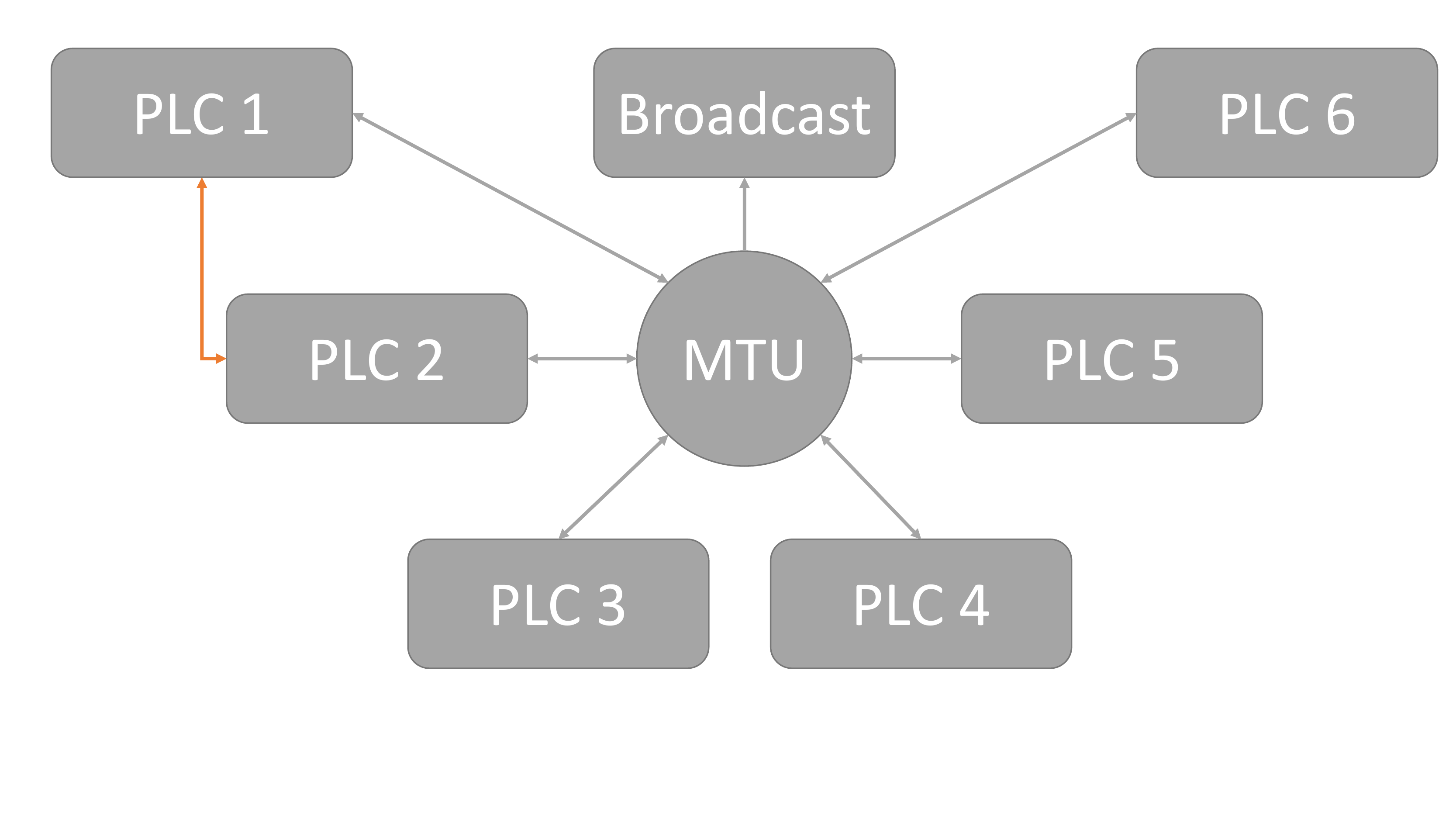}
\caption{Communication Structure of Data Set \textit{Send\_a\_fake\_command\_Modbus\_6RTU\_with\_operate}}
\label{fig:graph3}
\end{figure}

Three of the data sets presented by \textit{Lemay and Fernandez} contain labeled malicious traffic,
namely \textit{CnC\_uploading\_exe\_modbus\_6RTU\_with\_operate}, 
\textit{Moving\_two\_files\_Modbus\_6RTU} and \textit{Send\_a\_fake\_command\_Modbus\_6RTU\_with\_operate}.
Each of them has been monitored in a network consisting of six \acp{plc} and one \ac{mtu}.
Due to the nature of the introduced attack traffic,
each data set contains communication that is not present during normal behaviour.
If the communication was drawn as a graph,
with the \acp{plc} and the \ac{mtu} being the nodes and any communication creating an edge between the nodes,
the attacks can be distinguished as anomalous edges.
This behaviour is shown in Figures~\ref{fig:graph1} to~\ref{fig:graph3}. 
In Figures~\ref{fig:graph1} and~\ref{fig:graph2},
there are malicious packets addressed at the broadcast address.
Additionally,
there are malicious communication activities among the entities. 
This depicts the attempt to move laterally.
In every scenario,
\ac{plc} 1 is infected by a malware and performs different attempts to maliciously affect other entities.
Furthermore,
in the scenarios \textit{CnC\_uploading\_exe\_modbus\_6RTU\_with\_operate} and \textit{Moving\_two\_files\_Modbus\_6RTU},
the \ac{mtu} is infected and performs malicious activities. \\ \par
These kinds of attacks are relatively easy to detect,
e.g. by considering the fan-in and fan-out of nodes.
If it is considered as a time series that corresponds to the polling intervals,
malicious changes in the behaviour are detected effectively.

\section{The Hybrid Approach}
\label{sec:hybrid_approach}
As shown in the previous sections,
the success of an intrusion detection approach depends on the kind of attack.
If an attacker has obtained access to the control infrastructure of a process and does not try to move laterally,
it is hard to detect from a network perspective.
The attacker is in a position to alter parameters in a way that looks genuine.
However,
since the attacker tries to change the process behaviour,
considering process parameters can be used to detect attacks.
On the other hand,
there are attacks that do not alter the process behaviour itself.
Instead,
network resources are used to move,
extract or upload data.
Even though this does not directly impact the process behaviour,
this constitutes unwanted activity.
It can easily be detected by considering communication graphs. 
For each of the data sets analysed in this work,
only one of these characteristics was present and could be used for intrusion detection.
The work of \textit{Lemay and Fernandez} focuses on malicious network traffic that can be detected by alterations in the communication pattern~\cite{Lemay.2016}.
However,
the process remains unchanged.
In contrast,
the data set provided by \textit{iTrust,
Centre for Research in Cyber Security,
Singapore University of Technology and Design} only focuses on the impact of changes in parameters on the process.
Network-based lateral movement,
pivoting,
command and control or data exfiltration are not considered.\\ \par
In combining both approaches,
i.e. a time series analysis of process data and a communication pattern analysis,
a holistic overview of attacks can be derived.
Furthermore,
the source and destination of malicious activity can be derived.
The hybrid approach can be mapped on an aggregation model~\cite{Duque_Anton.2017b}.
As one of the most important requirements on industrial intrusion detection is functionality without feedback,
this needs to be addressed by any \ac{ids}.
Network information can be obtained from routers by mirroring the ports.
The traffic should normally not exceed the throughput limit of the mirror ports.
Host data,
such as process information on the \ac{hmi},
can be gathered there directly.
The integration and aggregation of all the data can be performed in a hierarchical fashion,
providing extensive information with no feedback,
e.g. in employing an aggregation concept as presented by \textit{Duque Anton et al.}~\cite{Duque_Anton.2017b}.
Furthermore,
the concept can be adapted to an online detection fashion.
This means that during operation,
a data set can be created containing topology and timing behaviour along nodes and edges.
Each newly introduced data point is compared to this data set,
calculating an anomaly score.
Since these methods do not need training as such,
set up is easily done.
Since the amount of data,
however,
is large in comparison to office \ac{it} networks,
storage and calculation will become tedious.
Some metric needs to be employed that summarises past events in order to efficiently calculate the anomaly score.

\section{Discussion}
\label{sec:discussion}
In this work,
two types of intrusion detection methods were presented and evaluated.
First,
a time series-based analysis method,
\textit{Matrix Profiles},
was used to detect attacks in the process parameters of an industrial process. 
The process is based on an industrial environment that performs water treatment in different steps.
Attacks in this work were introduced without consideration of the parameter breach,
thus leaving no trace in the network communication.
However,
the uniform and periodic nature of industrial processes enables the analysis method to detect the effects of any attack in an unsupervised manner.
Any outlier has been indicated by the application of \textit{Matrix Profiles}.
A major advantage is the small number of hyperparameters - one - as well as the ease of use.
Furthermore,
\textit{Matrix Profiles} do not have a distinct training phase, 
making employment and set-up feasible with low effort.
There are two main assumptions about outlier detection by distance calculation:
First,
any outlier is considered as an attack.
With this method,
there is no way to distinguish between misconfiguration and attacks.
Second,
attacks can only affect the process parameters in a unique fashion.
This means that if an attack has the exact same impact on the process a second time,
it will not be detected as an outlier by \textit{Matrix Profiles},
as one similar motif will then create a low minimal distance.
This can be mitigated by keeping track of the number of similar motifs in a time series.
Furthermore,
threshold calculation is a non-trivial challenge in unsupervised learning.
In the evaluation,
any attack could easily be detected by sight.
However,
formally setting a threshold after which an automated \ac{ids} triggers an alarm depends on the nature of the signal and the choice of $m$.\\ \par
Second, 
a different data set,
that has already been discussed in literature,
e.g. by \textit{Duque Anton et al.}~\cite{Duque_Anton.2018b, Duque_Anton.2018c},
has been analysed with respect to structure.
It can be seen that the attacks introduced in the data set change the topology of the communication,
even though they do not impact the process behaviour itself. \\ \par
Finally,
the concept for integrating both methods in order to detect various kinds of attacks,
based on meta-data as well as process information is presented.
It should address different types of attacks and detect an attack in different stages.
Usually,
after a perimeter breach,
lateral movement is attempted,
preceded by reconnaissance.
This creates anomalous traffic while not changing the process.
After pivoting,
an attacker could attempt to alter process parameters,
changing the behaviour.
If one of the stages is not detected by the \ac{ids},
detecting the attack in another stage is likely.
In addition to the presented ways of detecting anomalies,
attacker attribution and deception technologies,
as presented by \textit{Fraunholz et al.} can aid detection and identification of attackers~\cite{Fraunholz.2017d, Fraunholz.2017f}.

%
\begin{acks}
This work has been supported by the Federal Ministry of Education and Research of the Federal Republic of Germany (Foerderkennzeichen 16KIS0932, IUNO Insec).
One of the data sets used in this work has been provided by \textit{iTrust,
Centre for Research in Cyber Security,
Singapore University of Technology and Design.}
The authors alone are responsible for the content of the paper.
\end{acks}

%
\bibliographystyle{ACM-Reference-Format}
\bibliography{literature}


\begin{thebibliography}{51}


\ifx \showCODEN    \undefined \def \showCODEN     #1{\unskip}     \fi
\ifx \showDOI      \undefined \def \showDOI       #1{#1}\fi
\ifx \showISBNx    \undefined \def \showISBNx     #1{\unskip}     \fi
\ifx \showISBNxiii \undefined \def \showISBNxiii  #1{\unskip}     \fi
\ifx \showISSN     \undefined \def \showISSN      #1{\unskip}     \fi
\ifx \showLCCN     \undefined \def \showLCCN      #1{\unskip}     \fi
\ifx \shownote     \undefined \def \shownote      #1{#1}          \fi
\ifx \showarticletitle \undefined \def \showarticletitle #1{#1}   \fi
\ifx \showURL      \undefined \def \showURL       {\relax}        \fi
\providecommand\bibfield[2]{#2}
\providecommand\bibinfo[2]{#2}
\providecommand\natexlab[1]{#1}
\providecommand\showeprint[2][]{arXiv:#2}

\bibitem[\protect\citeauthoryear{Akoglu and Faloutsos}{Akoglu and
  Faloutsos}{2010}]%
        {Akoglu.2010b}
\bibfield{author}{\bibinfo{person}{Leman Akoglu} {and}
  \bibinfo{person}{Christos Faloutsos}.} \bibinfo{year}{2010}\natexlab{}.
\newblock \showarticletitle{Event Detection in Time Series of Mobile
  Communication Graphs}. In \bibinfo{booktitle}{\emph{Army Science
  Conference}}. \bibinfo{pages}{77--79}.
\newblock


\bibitem[\protect\citeauthoryear{Akoglu, Tong, and Koutra}{Akoglu
  et~al\mbox{.}}{2014}]%
        {Akoglu.2014}
\bibfield{author}{\bibinfo{person}{Leman Akoglu}, \bibinfo{person}{Hanghang
  Tong}, {and} \bibinfo{person}{Danai Koutra}.}
  \bibinfo{year}{2014}\natexlab{}.
\newblock \showarticletitle{Graph based Anomaly Detection and Description: A
  Survey}. In \bibinfo{booktitle}{\emph{Data Mining and Knowledge Discovery}},
  Vol.~\bibinfo{volume}{29}. \bibinfo{pages}{626--688}.
\newblock


\bibitem[\protect\citeauthoryear{Caselli, Zambon, and Kargl}{Caselli
  et~al\mbox{.}}{2015}]%
        {Caselli.2015}
\bibfield{author}{\bibinfo{person}{Marco Caselli}, \bibinfo{person}{Emmanuele
  Zambon}, {and} \bibinfo{person}{Frank Kargl}.}
  \bibinfo{year}{2015}\natexlab{}.
\newblock \showarticletitle{Sequence-aware Intrusion Detection in Industrial
  Control Systems}. In \bibinfo{booktitle}{\emph{Proceedings of the 1st ACM
  Workshop on Cyber-Physical System Security}} \emph{(\bibinfo{series}{CPSS
  '15})}. \bibinfo{publisher}{ACM}, \bibinfo{address}{New York, NY, USA},
  \bibinfo{pages}{13--24}.
\newblock
\showISBNx{978-1-4503-3448-8}
\urldef\tempurl%
\url{https://doi.org/10.1145/2732198.2732200}
\showDOI{\tempurl}


\bibitem[\protect\citeauthoryear{Cheung, Crawford, Dilger, Frank, Hoagland,
  Levitt, Wee, Staniford-Chen, Yip, and Zerkle}{Cheung et~al\mbox{.}}{1996}]%
        {Cheung.1999}
\bibfield{author}{\bibinfo{person}{Steven Cheung}, \bibinfo{person}{Rick
  Crawford}, \bibinfo{person}{Mark Dilger}, \bibinfo{person}{Jeremy Frank},
  \bibinfo{person}{Jim Hoagland}, \bibinfo{person}{Karl Levitt},
  \bibinfo{person}{C. Wee}, \bibinfo{person}{Stuart Staniford-Chen},
  \bibinfo{person}{Raymond Yip}, {and} \bibinfo{person}{Dan Zerkle}.}
  \bibinfo{year}{1996}\natexlab{}.
\newblock \bibinfo{booktitle}{\emph{The Design of {GrIDS}: A Graph Based
  Intrusion Detection System for Large Networks}}.
\newblock \bibinfo{type}{{CSE-99-2}}. \bibinfo{institution}{UC Davis Computer
  Science Department}.
\newblock


\bibitem[\protect\citeauthoryear{Dau and Keogh}{Dau and Keogh}{2017}]%
        {Dau.2017}
\bibfield{author}{\bibinfo{person}{Hoang~Anh Dau} {and} \bibinfo{person}{Eamonn
  Keogh}.} \bibinfo{year}{2017}\natexlab{}.
\newblock \showarticletitle{{Matrix Profile V}: A Generic Technique to
  Incorporate Domain Knowledge into Motif Discovery}. In
  \bibinfo{booktitle}{\emph{Proceedings of the 23rd ACM SIGKDD International
  Conference on Knowledge Discovery and Data Mining}}
  \emph{(\bibinfo{series}{KDD '17})}. \bibinfo{publisher}{ACM},
  \bibinfo{address}{New York, NY, USA}, \bibinfo{pages}{125--134}.
\newblock
\showISBNx{978-1-4503-4887-4}
\urldef\tempurl%
\url{https://doi.org/10.1145/3097983.3097993}
\showDOI{\tempurl}


\bibitem[\protect\citeauthoryear{Debar, Becker, and Siboni}{Debar
  et~al\mbox{.}}{1992}]%
        {Debar.1992}
\bibfield{author}{\bibinfo{person}{Herve Debar}, \bibinfo{person}{Monique
  Becker}, {and} \bibinfo{person}{Didier Siboni}.}
  \bibinfo{year}{1992}\natexlab{}.
\newblock \showarticletitle{A Neural Network Component for an Intrusion
  Detection System}. In \bibinfo{booktitle}{\emph{IEEE Symposium on Security
  and Privacy}}. \bibinfo{pages}{240--250}.
\newblock


\bibitem[\protect\citeauthoryear{Dethlefs}{Dethlefs}{2015}]%
        {Dethlefs.2015}
\bibfield{author}{\bibinfo{person}{Robert Dethlefs}.}
  \bibinfo{year}{2015}\natexlab{}.
\newblock \showarticletitle{How cyber attacks became more profitable than the
  drug trade}.
\newblock \bibinfo{journal}{\emph{Fortune}} (\bibinfo{year}{2015}).
\newblock


\bibitem[\protect\citeauthoryear{{Duque Anton}, Ahrens, Fraunholz, and
  Schotten}{{Duque Anton} et~al\mbox{.}}{2018a}]%
        {Duque_Anton.2018c}
\bibfield{author}{\bibinfo{person}{Simon {Duque Anton}}, \bibinfo{person}{Lia
  Ahrens}, \bibinfo{person}{Daniel Fraunholz}, {and}
  \bibinfo{person}{Hans~Dieter Schotten}.} \bibinfo{year}{2018}\natexlab{a}.
\newblock \showarticletitle{Time is of the Essence: Machine Learning-based
  Intrusion Detection in Industrial Time Series Data}. In
  \bibinfo{booktitle}{\emph{Proceedings of the 2018 IEEE International
  Conference on Data Mining Workshops (ICDMW)}}. \bibinfo{publisher}{IEEE}.
\newblock


\bibitem[\protect\citeauthoryear{Duque~Anton, Fraunholz, Lipps, Pohl,
  Zimmermann, and Schotten}{Duque~Anton et~al\mbox{.}}{2017a}]%
        {Duque_Anton.2017a}
\bibfield{author}{\bibinfo{person}{Simon Duque~Anton}, \bibinfo{person}{Daniel
  Fraunholz}, \bibinfo{person}{Christoph Lipps}, \bibinfo{person}{Frederic
  Pohl}, \bibinfo{person}{Marc Zimmermann}, {and} \bibinfo{person}{Hans~Dieter
  Schotten}.} \bibinfo{year}{2017}\natexlab{a}.
\newblock \showarticletitle{Two Decades of {SCADA} Exploitation: A Brief
  History}. In \bibinfo{booktitle}{\emph{2017 IEEE Conference on Application,
  Information and Network Security (AINS)}}. \bibinfo{pages}{98--104}.
\newblock
\urldef\tempurl%
\url{https://doi.org/10.1109/AINS.2017.8270432}
\showDOI{\tempurl}


\bibitem[\protect\citeauthoryear{Duque~Anton, Fraunholz, Zemitis, Pohl, and
  Schotten}{Duque~Anton et~al\mbox{.}}{2017b}]%
        {Duque_Anton.2017b}
\bibfield{author}{\bibinfo{person}{Simon Duque~Anton}, \bibinfo{person}{Daniel
  Fraunholz}, \bibinfo{person}{Janis Zemitis}, \bibinfo{person}{Frederic Pohl},
  {and} \bibinfo{person}{Hans~Dieter Schotten}.}
  \bibinfo{year}{2017}\natexlab{b}.
\newblock \showarticletitle{Highly Scalable and Flexible Model for Effective
  Aggregation of Context-based Data in Generic IIoT Scenarios}. In
  \bibinfo{booktitle}{\emph{9th Central European Workshop on Services and their
  Composition (ZEUS-2017), February 13-14, Lugano, Switzerland}}.
  \bibinfo{pages}{51--58}.
\newblock


\bibitem[\protect\citeauthoryear{{Duque Anton}, Kanoor, Fraunholz, and
  Schotten}{{Duque Anton} et~al\mbox{.}}{2018b}]%
        {Duque_Anton.2018b}
\bibfield{author}{\bibinfo{person}{Simon {Duque Anton}},
  \bibinfo{person}{Suneetha Kanoor}, \bibinfo{person}{Daniel Fraunholz}, {and}
  \bibinfo{person}{Hans~Dieter Schotten}.} \bibinfo{year}{2018}\natexlab{b}.
\newblock \showarticletitle{Evaluation of Machine Learning-based Anomaly
  Detection Algorithms on an Industrial Modbus/TCP Data Set}. In
  \bibinfo{booktitle}{\emph{Proceedings of the 13th International Conference on
  Availability, Reliability and Security (ARES)}}. \bibinfo{publisher}{ACM}.
\newblock


\bibitem[\protect\citeauthoryear{Eberle and Holder}{Eberle and Holder}{2007}]%
        {Eberle.2007}
\bibfield{author}{\bibinfo{person}{WIlliam Eberle} {and}
  \bibinfo{person}{Lawrence Holder}.} \bibinfo{year}{2007}\natexlab{}.
\newblock \showarticletitle{Discovering Structural Anomalies in Graph-Based
  Data}. In \bibinfo{booktitle}{\emph{Seventh IEEE International Conference on
  Data Mining Workshops (ICDMW 2007)}}. \bibinfo{pages}{393--398}.
\newblock


\bibitem[\protect\citeauthoryear{Eswaran and Faloutsos}{Eswaran and
  Faloutsos}{2018}]%
        {Eswaran.2018}
\bibfield{author}{\bibinfo{person}{Dhivya Eswaran} {and}
  \bibinfo{person}{Christos Faloutsos}.} \bibinfo{year}{2018}\natexlab{}.
\newblock \showarticletitle{SedanSpot: Detecting Anomalies in Edge Streams}. In
  \bibinfo{booktitle}{\emph{2018 IEEE International Conference on Data Mining
  (ICDM)}}. \bibinfo{pages}{953--958}.
\newblock
\showISSN{2374-8486}
\urldef\tempurl%
\url{https://doi.org/10.1109/ICDM.2018.00117}
\showDOI{\tempurl}


\bibitem[\protect\citeauthoryear{Ferdousi and Maeda}{Ferdousi and
  Maeda}{2006}]%
        {Ferdousi.2006}
\bibfield{author}{\bibinfo{person}{Z. Ferdousi} {and} \bibinfo{person}{A.
  Maeda}.} \bibinfo{year}{2006}\natexlab{}.
\newblock \showarticletitle{Unsupervised Outlier Detection in Time Series
  Data}. In \bibinfo{booktitle}{\emph{22nd International Conference on Data
  Engineering Workshops (ICDEW'06)}}.
\newblock
\urldef\tempurl%
\url{https://doi.org/10.1109/ICDEW.2006.157}
\showDOI{\tempurl}


\bibitem[\protect\citeauthoryear{Fovino, Carcano, De~Lacheze~Murel, Trombetta,
  and Masera}{Fovino et~al\mbox{.}}{2010}]%
        {Fovino.2010}
\bibfield{author}{\bibinfo{person}{Igor~Nai Fovino}, \bibinfo{person}{Andrea
  Carcano}, \bibinfo{person}{Thibault De~Lacheze~Murel},
  \bibinfo{person}{Alberto Trombetta}, {and} \bibinfo{person}{Marcelo Masera}.}
  \bibinfo{year}{2010}\natexlab{}.
\newblock \showarticletitle{{Modbus/DNP3} State-Based Intrusion Detection
  System}. In \bibinfo{booktitle}{\emph{24th IEEE International Conference on
  Advanced Information Networking and Applications(AINA)}}.
  \bibinfo{pages}{729--736}.
\newblock


\bibitem[\protect\citeauthoryear{Fraunholz, {Duque Anton}, and
  Schotten}{Fraunholz et~al\mbox{.}}{2017a}]%
        {Fraunholz.2017d}
\bibfield{author}{\bibinfo{person}{Daniel Fraunholz}, \bibinfo{person}{Simon
  {Duque Anton}}, {and} \bibinfo{person}{Hans~Dieter Schotten}.}
  \bibinfo{year}{2017}\natexlab{a}.
\newblock \showarticletitle{Introducing GAMfIS: A Generic Attacker Model for
  Information Security}.
\newblock \bibinfo{journal}{\emph{International Conference on Software,
  Telecommunications and Computer Networks}}  \bibinfo{volume}{25}
  (\bibinfo{year}{2017}).
\newblock


\bibitem[\protect\citeauthoryear{Fraunholz, Krohmer, {Duque Anton}, and
  Schotten}{Fraunholz et~al\mbox{.}}{2017b}]%
        {Fraunholz.2017f}
\bibfield{author}{\bibinfo{person}{Daniel Fraunholz}, \bibinfo{person}{Daniel
  Krohmer}, \bibinfo{person}{Simon {Duque Anton}}, {and}
  \bibinfo{person}{Hans~Dieter Schotten}.} \bibinfo{year}{2017}\natexlab{b}.
\newblock \showarticletitle{YAAS - On the Attribution of Honeypot Data}.
\newblock \bibinfo{journal}{\emph{International Journal on Cyber Situational
  Awareness}} \bibinfo{volume}{2}, \bibinfo{number}{1} (\bibinfo{year}{2017}),
  \bibinfo{pages}{31--48}.
\newblock


\bibitem[\protect\citeauthoryear{Gao and Morris}{Gao and Morris}{2014}]%
        {Gao.2014}
\bibfield{author}{\bibinfo{person}{Wei Gao} {and} \bibinfo{person}{Thomas~H.
  Morris}.} \bibinfo{year}{2014}\natexlab{}.
\newblock \showarticletitle{On Cyber Attacks and Signature Based Intrusion
  Detection for Modbus Based Industrial Control Systems}.
\newblock \bibinfo{journal}{\emph{Journal of Digital Forensics, Security and
  Law}} \bibinfo{volume}{9}, \bibinfo{number}{1} (\bibinfo{year}{2014}).
\newblock


\bibitem[\protect\citeauthoryear{Garcia-Teodoro, Diaz-Verdejo, Macia-Fernandez,
  and Vazquez}{Garcia-Teodoro et~al\mbox{.}}{2008}]%
        {Garcia-Teodoro.2008}
\bibfield{author}{\bibinfo{person}{P. Garcia-Teodoro}, \bibinfo{person}{J.
  Diaz-Verdejo}, \bibinfo{person}{G. Macia-Fernandez}, {and}
  \bibinfo{person}{E. Vazquez}.} \bibinfo{year}{2008}\natexlab{}.
\newblock \showarticletitle{Anomaly-based network intrusion detection:
  Techniques, systems and challenges}.
\newblock \bibinfo{journal}{\emph{Computers {\&} Security}}
  \bibinfo{volume}{28}, \bibinfo{number}{1-2} (\bibinfo{date}{August}
  \bibinfo{year}{2008}), \bibinfo{pages}{18--28}.
\newblock


\bibitem[\protect\citeauthoryear{Ghaeini and Tippenhauer}{Ghaeini and
  Tippenhauer}{2016}]%
        {Ghaeini.2016}
\bibfield{author}{\bibinfo{person}{Hamid~Reza Ghaeini} {and}
  \bibinfo{person}{Nils~Ole Tippenhauer}.} \bibinfo{year}{2016}\natexlab{}.
\newblock \showarticletitle{{HAMIDS}: Hierarchical Monitoring Intrusion
  Detection System for Industrial Control Systems}. In
  \bibinfo{booktitle}{\emph{Proceedings of the 2nd ACM Workshop on
  Cyber-Physical Systems Security and Privacy}} \emph{(\bibinfo{series}{CPS-SPC
  '16})}. \bibinfo{publisher}{ACM}, \bibinfo{address}{New York, NY, USA},
  \bibinfo{pages}{103--111}.
\newblock
\showISBNx{978-1-4503-4568-2}
\urldef\tempurl%
\url{https://doi.org/10.1145/2994487.2994492}
\showDOI{\tempurl}


\bibitem[\protect\citeauthoryear{Goh, Adepu, Junejo, and Mathur}{Goh
  et~al\mbox{.}}{2016}]%
        {Goh.2016}
\bibfield{author}{\bibinfo{person}{Jonathan Goh}, \bibinfo{person}{Sridhar
  Adepu}, \bibinfo{person}{Khurum~Nazir Junejo}, {and} \bibinfo{person}{Aditya
  Mathur}.} \bibinfo{year}{2016}\natexlab{}.
\newblock \showarticletitle{A Dataset to Support Research in the Design of
  Secure Water Treatment Systems}. In \bibinfo{booktitle}{\emph{Proceedings of
  the 11th International Conference on Critical Information Infrastructures
  Security}}.
\newblock


\bibitem[\protect\citeauthoryear{Greenberg}{Greenberg}{2017}]%
        {Greenberg.2017}
\bibfield{author}{\bibinfo{person}{Andy Greenberg}.}
  \bibinfo{year}{2017}\natexlab{}.
\newblock \showarticletitle{'Crash Override': The Malware that Took Down a
  Power Grid}.
\newblock \bibinfo{journal}{\emph{Wired}} (\bibinfo{year}{2017}).
\newblock


\bibitem[\protect\citeauthoryear{Hadeli, Schierholz, Braendle, and
  Tuduce}{Hadeli et~al\mbox{.}}{2009}]%
        {Hadeli.2009}
\bibfield{author}{\bibinfo{person}{Hadeli Hadeli}, \bibinfo{person}{Ragnar
  Schierholz}, \bibinfo{person}{Markus Braendle}, {and}
  \bibinfo{person}{Cristian Tuduce}.} \bibinfo{year}{2009}\natexlab{}.
\newblock \showarticletitle{Leveraging determinism in industrial control
  systems for advanced anomaly detection and reliable security configuration}.
  In \bibinfo{booktitle}{\emph{2009 IEEE Conference on Emerging Technologies
  Factory Automation}}. \bibinfo{pages}{1--8}.
\newblock
\showISSN{1946-0740}
\urldef\tempurl%
\url{https://doi.org/10.1109/ETFA.2009.5347134}
\showDOI{\tempurl}


\bibitem[\protect\citeauthoryear{Haller, Karnouskos, and Schroth}{Haller
  et~al\mbox{.}}{2008}]%
        {Haller.2008}
\bibfield{author}{\bibinfo{person}{Stephan Haller}, \bibinfo{person}{Stamatis
  Karnouskos}, {and} \bibinfo{person}{Christoph Schroth}.}
  \bibinfo{year}{2008}\natexlab{}.
\newblock \showarticletitle{The Internet of Things in an Enterprise Context}.
  In \bibinfo{booktitle}{\emph{Future Internet Symposium}}.
  \bibinfo{publisher}{Springer-Verlag, Berlin, Heidelberg},
  \bibinfo{pages}{14--28}.
\newblock
\urldef\tempurl%
\url{https://doi.org/10.1007/978-3-642-00985-3\_2}
\showDOI{\tempurl}


\bibitem[\protect\citeauthoryear{Hamdi and Boudriga}{Hamdi and
  Boudriga}{2009}]%
        {Hamdi.2007}
\bibfield{author}{\bibinfo{person}{Mohamed Hamdi} {and}
  \bibinfo{person}{Noureddine Boudriga}.} \bibinfo{year}{2009}\natexlab{}.
\newblock \showarticletitle{Detecting Denial-of-Service attacks using the
  wavelet transform}.
\newblock \bibinfo{journal}{\emph{Computer Communications}}
  \bibinfo{volume}{30}, \bibinfo{number}{16} (\bibinfo{date}{November}
  \bibinfo{year}{2009}).
\newblock
\urldef\tempurl%
\url{https://doi.org/10.1016/j.comcom.2007.05.061}
\showDOI{\tempurl}


\bibitem[\protect\citeauthoryear{{iTrust Centre for Research in Cyber
  Security}}{{iTrust Centre for Research in Cyber Security}}{2018}]%
        {iTrust.2018}
\bibfield{author}{\bibinfo{person}{{iTrust Centre for Research in Cyber
  Security}}.} \bibinfo{year}{2018}\natexlab{}.
\newblock \bibinfo{booktitle}{\emph{Secure Water Treatment {(SWaT)} Testbed}}.
\newblock \bibinfo{type}{{T}echnical {R}eport} 4.2.
  \bibinfo{institution}{Singapore University of Technology and Design}.
\newblock


\bibitem[\protect\citeauthoryear{Jyothsna and Rama~Prasad}{Jyothsna and
  Rama~Prasad}{2011}]%
        {Jyothsna.2011}
\bibfield{author}{\bibinfo{person}{V. Jyothsna} {and} \bibinfo{person}{V.~V.
  Rama~Prasad}.} \bibinfo{year}{2011}\natexlab{}.
\newblock \showarticletitle{A Review of Anomaly based IntrusionDetection
  Systems}.
\newblock \bibinfo{journal}{\emph{International Journal of Computer
  Applications}} \bibinfo{volume}{28}, \bibinfo{number}{7}
  (\bibinfo{date}{September} \bibinfo{year}{2011}), \bibinfo{pages}{26--35}.
\newblock


\bibitem[\protect\citeauthoryear{Khalili and Sami}{Khalili and Sami}{2015}]%
        {Khalili.2015}
\bibfield{author}{\bibinfo{person}{Abdullah Khalili} {and}
  \bibinfo{person}{Ashkan Sami}.} \bibinfo{year}{2015}\natexlab{}.
\newblock \showarticletitle{{SysDetect}: A systematic approach to critical
  state determination for Industrial Intrusion Detection Systems using Apriori
  algorithm}.
\newblock \bibinfo{journal}{\emph{Journal of Process Control}}
  \bibinfo{volume}{32} (\bibinfo{date}{January} \bibinfo{year}{2015}),
  \bibinfo{pages}{154--160}.
\newblock
\urldef\tempurl%
\url{https://doi.org/10.1016/j.jprocont.2015.04.005}
\showDOI{\tempurl}


\bibitem[\protect\citeauthoryear{Langner}{Langner}{2013}]%
        {Langner.2013}
\bibfield{author}{\bibinfo{person}{Ralph Langner}.}
  \bibinfo{year}{2013}\natexlab{}.
\newblock \bibinfo{booktitle}{\emph{To Kill a Centrifuge}}.
\newblock \bibinfo{type}{{T}echnical {R}eport}. \bibinfo{institution}{The
  Langner Group}.
\newblock


\bibitem[\protect\citeauthoryear{Lemay and Fernandez}{Lemay and
  Fernandez}{2016}]%
        {Lemay.2016}
\bibfield{author}{\bibinfo{person}{Antoine Lemay} {and}
  \bibinfo{person}{Jose~M. Fernandez}.} \bibinfo{year}{2016}\natexlab{}.
\newblock \showarticletitle{Providing {SCADA} Network Data Sets for Intrusion
  Detection Research}. In \bibinfo{booktitle}{\emph{9th Workshop on Cyber
  Security Experimentation and Test (CSET 16)}}. \bibinfo{address}{Austin, TX}.
\newblock


\bibitem[\protect\citeauthoryear{Lu and Ghorbani}{Lu and Ghorbani}{2009}]%
        {Lu.2009}
\bibfield{author}{\bibinfo{person}{Wei Lu} {and} \bibinfo{person}{Ali~A.
  Ghorbani}.} \bibinfo{year}{2009}\natexlab{}.
\newblock \showarticletitle{Network Anomaly Detection Based on Wavelet
  Analysis}.
\newblock \bibinfo{journal}{\emph{EURASIP J. Adv. Signal Process}}
  \bibinfo{volume}{2009}, Article \bibinfo{articleno}{4}
  (\bibinfo{date}{January} \bibinfo{year}{2009}), \bibinfo{numpages}{16}~pages.
\newblock
\showISSN{1110-8657}
\urldef\tempurl%
\url{https://doi.org/10.1155/2009/837601}
\showDOI{\tempurl}


\bibitem[\protect\citeauthoryear{Ma and Perkins}{Ma and Perkins}{2003}]%
        {Ma.2003}
\bibfield{author}{\bibinfo{person}{J. Ma} {and} \bibinfo{person}{S. Perkins}.}
  \bibinfo{year}{2003}\natexlab{}.
\newblock \showarticletitle{Time-series novelty detection using one-class
  support vector machines}. In \bibinfo{booktitle}{\emph{Proceedings of the
  International Joint Conference on Neural Networks}},
  Vol.~\bibinfo{volume}{3}. \bibinfo{pages}{1741--1745}.
\newblock
\showISSN{1098-7576}
\urldef\tempurl%
\url{https://doi.org/10.1109/IJCNN.2003.1223670}
\showDOI{\tempurl}


\bibitem[\protect\citeauthoryear{Moayedi and Masnadi-Shirazi}{Moayedi and
  Masnadi-Shirazi}{2008}]%
        {Moayedi.2008}
\bibfield{author}{\bibinfo{person}{H.~Zare Moayedi} {and}
  \bibinfo{person}{M.~A. Masnadi-Shirazi}.} \bibinfo{year}{2008}\natexlab{}.
\newblock \showarticletitle{Arima model for network traffic prediction and
  anomaly detection}. In \bibinfo{booktitle}{\emph{2008 International Symposium
  on Information Technology}}, Vol.~\bibinfo{volume}{4}. \bibinfo{pages}{1--6}.
\newblock
\showISSN{2155-8973}
\urldef\tempurl%
\url{https://doi.org/10.1109/ITSIM.2008.4631947}
\showDOI{\tempurl}


\bibitem[\protect\citeauthoryear{Morris, Vaughn, and Dandass}{Morris
  et~al\mbox{.}}{2012}]%
        {Morris.2012}
\bibfield{author}{\bibinfo{person}{Thomas Morris}, \bibinfo{person}{Rayford
  Vaughn}, {and} \bibinfo{person}{Yoginder Dandass}.}
  \bibinfo{year}{2012}\natexlab{}.
\newblock \showarticletitle{A Retrofit Network Intrusion Detection System for
  {MODBUS RTU and ASCII} Industrial Control Systems}. In
  \bibinfo{booktitle}{\emph{2012 45th Hawaii International Conference on System
  Sciences}}. \bibinfo{pages}{2338--2345}.
\newblock
\showISSN{1530-1605}
\urldef\tempurl%
\url{https://doi.org/10.1109/HICSS.2012.78}
\showDOI{\tempurl}


\bibitem[\protect\citeauthoryear{Munz and Carle}{Munz and Carle}{2007}]%
        {Munz.2007}
\bibfield{author}{\bibinfo{person}{Gerhard Munz} {and} \bibinfo{person}{Georg
  Carle}.} \bibinfo{year}{2007}\natexlab{}.
\newblock \showarticletitle{Real-time Analysis of Flow Data for Network Attack
  Detection}. In \bibinfo{booktitle}{\emph{2007 10th IFIP/IEEE International
  Symposium on Integrated Network Management}}. \bibinfo{pages}{100--108}.
\newblock
\showISSN{1573-0077}
\urldef\tempurl%
\url{https://doi.org/10.1109/INM.2007.374774}
\showDOI{\tempurl}


\bibitem[\protect\citeauthoryear{Noble and Cook}{Noble and Cook}{2003}]%
        {Noble.2003}
\bibfield{author}{\bibinfo{person}{Caleb~C. Noble} {and}
  \bibinfo{person}{Diane~J. Cook}.} \bibinfo{year}{2003}\natexlab{}.
\newblock \showarticletitle{Graph-based Anomaly Detection}. In
  \bibinfo{booktitle}{\emph{Proceedings of the Ninth ACM SIGKDD International
  Conference on Knowledge Discovery and Data Mining}}
  \emph{(\bibinfo{series}{KDD '03})}. \bibinfo{publisher}{ACM},
  \bibinfo{address}{New York, NY, USA}, \bibinfo{pages}{631--636}.
\newblock
\showISBNx{1-58113-737-0}
\urldef\tempurl%
\url{https://doi.org/10.1145/956750.956831}
\showDOI{\tempurl}


\bibitem[\protect\citeauthoryear{Pasqualetti, Doerfler, and Bullo}{Pasqualetti
  et~al\mbox{.}}{2013}]%
        {Pasqualetti.2013}
\bibfield{author}{\bibinfo{person}{Fabio Pasqualetti}, \bibinfo{person}{Florian
  Doerfler}, {and} \bibinfo{person}{Franceso Bullo}.}
  \bibinfo{year}{2013}\natexlab{}.
\newblock \showarticletitle{Attack Detection and Identification in
  Cyber-Physical Systems}.
\newblock \bibinfo{journal}{\emph{IEEE Trans. Automat. Control}}
  \bibinfo{volume}{58}, \bibinfo{number}{11} (\bibinfo{date}{November}
  \bibinfo{year}{2013}), \bibinfo{pages}{2715--2729}.
\newblock
\showISSN{0018-9286}
\urldef\tempurl%
\url{https://doi.org/10.1109/TAC.2013.2266831}
\showDOI{\tempurl}


\bibitem[\protect\citeauthoryear{Ponomarev and Atkison}{Ponomarev and
  Atkison}{2016}]%
        {Ponomarev.2016}
\bibfield{author}{\bibinfo{person}{Stanislav Ponomarev} {and}
  \bibinfo{person}{Travis Atkison}.} \bibinfo{year}{2016}\natexlab{}.
\newblock \showarticletitle{Industrial Control System Network Intrusion
  Detection by Telemetry Analysis}.
\newblock \bibinfo{journal}{\emph{IEEE Transactions on Dependable and Secure
  Computing}} \bibinfo{volume}{13}, \bibinfo{number}{2} (\bibinfo{date}{March}
  \bibinfo{year}{2016}), \bibinfo{pages}{252--260}.
\newblock
\showISSN{1545-5971}
\urldef\tempurl%
\url{https://doi.org/10.1109/TDSC.2015.2443793}
\showDOI{\tempurl}


\bibitem[\protect\citeauthoryear{Regis~Barbosa and Pras}{Regis~Barbosa and
  Pras}{2010}]%
        {Regis_Barbosa.2010}
\bibfield{author}{\bibinfo{person}{Rafael~Ramos Regis~Barbosa} {and}
  \bibinfo{person}{Aiko Pras}.} \bibinfo{year}{2010}\natexlab{}.
\newblock \showarticletitle{Intrusion Detection in {SCADA} Networks}.
\newblock \bibinfo{journal}{\emph{Mechanisms for Autonomous Management of
  Networks and Services}}  \bibinfo{volume}{6155} (\bibinfo{year}{2010}).
\newblock
\urldef\tempurl%
\url{https://doi.org/10.1007/978-3-642-13986-4\_23}
\showDOI{\tempurl}


\bibitem[\protect\citeauthoryear{Schneider and B\"{o}ttinger}{Schneider and
  B\"{o}ttinger}{2018}]%
        {Schneider.2018}
\bibfield{author}{\bibinfo{person}{Peter Schneider} {and}
  \bibinfo{person}{Konstantin B\"{o}ttinger}.} \bibinfo{year}{2018}\natexlab{}.
\newblock \showarticletitle{High-Performance Unsupervised Anomaly Detection for
  Cyber-Physical System Networks}. In \bibinfo{booktitle}{\emph{Proceedings of
  the 2018 Workshop on Cyber-Physical Systems Security and PrivaCy}}
  \emph{(\bibinfo{series}{CPS-SPC '18})}. \bibinfo{publisher}{ACM},
  \bibinfo{address}{New York, NY, USA}, \bibinfo{pages}{1--12}.
\newblock
\showISBNx{978-1-4503-5992-4}
\urldef\tempurl%
\url{https://doi.org/10.1145/3264888.3264890}
\showDOI{\tempurl}


\bibitem[\protect\citeauthoryear{Staniford-Chen, Cheung, Crawford, Dilger,
  Frank, Hoagland, Levitt, Wee, Yip, and Zerkle}{Staniford-Chen
  et~al\mbox{.}}{1996}]%
        {Staniford-Chen.1996}
\bibfield{author}{\bibinfo{person}{Stuart Staniford-Chen},
  \bibinfo{person}{Steven Cheung}, \bibinfo{person}{Rick Crawford},
  \bibinfo{person}{Mark Dilger}, \bibinfo{person}{Jeremy Frank},
  \bibinfo{person}{Jim Hoagland}, \bibinfo{person}{Karl Levitt},
  \bibinfo{person}{C. Wee}, \bibinfo{person}{Raymond Yip}, {and}
  \bibinfo{person}{Dan Zerkle}.} \bibinfo{year}{1996}\natexlab{}.
\newblock \showarticletitle{{GrIDS} - A Graph Based Intrusion Detection System
  for Large Networks}. In \bibinfo{booktitle}{\emph{Proceedings of the 19th
  National Information Systems Security Conference}}, Vol.~\bibinfo{volume}{1}.
  \bibinfo{pages}{361--370}.
\newblock


\bibitem[\protect\citeauthoryear{Swiler and Phillips}{Swiler and
  Phillips}{1998}]%
        {Swiler.1998}
\bibfield{author}{\bibinfo{person}{Laura~Painton Swiler} {and}
  \bibinfo{person}{Cynthia Phillips}.} \bibinfo{year}{1998}\natexlab{}.
\newblock \showarticletitle{A Graph-Based System for Network-Vulnerability
  Analysis}.
\newblock  (\bibinfo{date}{June} \bibinfo{year}{1998}).
\newblock
\urldef\tempurl%
\url{https://doi.org/10.2172/573291}
\showDOI{\tempurl}


\bibitem[\protect\citeauthoryear{Symantec}{Symantec}{2009}]%
        {Symantec.2009}
\bibfield{author}{\bibinfo{person}{Symantec}.} \bibinfo{year}{2009}\natexlab{}.
\newblock \bibinfo{booktitle}{\emph{Cyber Crime has Surpassed Illegal Drug
  Trafficking as a Criminal Moneymaker; 1 in 5 will become a Victim.}}
\newblock
\urldef\tempurl%
\url{https://www.symantec.com/about/newsroom/press-releases/2009/symantec\_0910\_01}
\showURL{%
\tempurl}


\bibitem[\protect\citeauthoryear{Tabatabaie~Nezhad, Nazari, and
  Gharavol}{Tabatabaie~Nezhad et~al\mbox{.}}{2016}]%
        {Tabatabaie.2016}
\bibfield{author}{\bibinfo{person}{Seyyed~Meysam Tabatabaie~Nezhad},
  \bibinfo{person}{Mahboubeh Nazari}, {and} \bibinfo{person}{Ebrahim~A.
  Gharavol}.} \bibinfo{year}{2016}\natexlab{}.
\newblock \showarticletitle{A Novel {DoS} and {DDoS} Attacks Detection
  Algorithm Using {ARIMA} Time Series Model and Chaotic System in Computer
  Networks}.
\newblock \bibinfo{journal}{\emph{IEEE Communications Letters}}
  \bibinfo{volume}{20}, \bibinfo{number}{4} (\bibinfo{date}{April}
  \bibinfo{year}{2016}), \bibinfo{pages}{700--703}.
\newblock
\showISSN{1089-7798}
\urldef\tempurl%
\url{https://doi.org/10.1109/LCOMM.2016.2517622}
\showDOI{\tempurl}


\bibitem[\protect\citeauthoryear{Tao, Hui, and Xiong}{Tao
  et~al\mbox{.}}{2018}]%
        {Tao.2018}
\bibfield{author}{\bibinfo{person}{Jialing Tao}, \bibinfo{person}{Wang Hui},
  {and} \bibinfo{person}{Tao Xiong}.} \bibinfo{year}{2018}\natexlab{}.
\newblock \showarticletitle{Selective Graph Attention Networks for Account
  Takeover Detection}. In \bibinfo{booktitle}{\emph{2018 IEEE International
  Conference on Data Mining Workshops (ICDMW)}}.
\newblock


\bibitem[\protect\citeauthoryear{Tsang and Kwong}{Tsang and Kwong}{2005}]%
        {Tsang.2005}
\bibfield{author}{\bibinfo{person}{Chi-Ho Tsang} {and} \bibinfo{person}{S.
  Kwong}.} \bibinfo{year}{2005}\natexlab{}.
\newblock \showarticletitle{Multi-agent intrusion detection system in
  industrial network using ant colony clustering approach and unsupervised
  feature extraction}. In \bibinfo{booktitle}{\emph{2005 IEEE International
  Conference on Industrial Technology}}. \bibinfo{pages}{51--56}.
\newblock
\urldef\tempurl%
\url{https://doi.org/10.1109/ICIT.2005.1600609}
\showDOI{\tempurl}


\bibitem[\protect\citeauthoryear{Uckelmann, Harrison, and
  Michahelles}{Uckelmann et~al\mbox{.}}{2011}]%
        {Uckelmann.2011}
\bibfield{author}{\bibinfo{person}{Dieter Uckelmann}, \bibinfo{person}{Mark
  Harrison}, {and} \bibinfo{person}{Florian Michahelles}.}
  \bibinfo{year}{2011}\natexlab{}.
\newblock \showarticletitle{An Architectural Approach Towards the Future
  Internet of Things}. In \bibinfo{booktitle}{\emph{Architecting the Internet
  of Things}}. \bibinfo{publisher}{Springer-Verlag, Berlin, Heidelberg},
  \bibinfo{pages}{1--24}.
\newblock
\urldef\tempurl%
\url{https://doi.org/10.1007/978-3-642-19157-2\_1}
\showDOI{\tempurl}


\bibitem[\protect\citeauthoryear{Yaacob, Tan, Chien, and Tan}{Yaacob
  et~al\mbox{.}}{2010}]%
        {Yaacob.2010}
\bibfield{author}{\bibinfo{person}{A.~H. Yaacob}, \bibinfo{person}{I.~K.~T.
  Tan}, \bibinfo{person}{S.~F. Chien}, {and} \bibinfo{person}{H.~K. Tan}.}
  \bibinfo{year}{2010}\natexlab{}.
\newblock \showarticletitle{ARIMA Based Network Anomaly Detection}. In
  \bibinfo{booktitle}{\emph{2010 Second International Conference on
  Communication Software and Networks}}. \bibinfo{pages}{205--209}.
\newblock
\urldef\tempurl%
\url{https://doi.org/10.1109/ICCSN.2010.55}
\showDOI{\tempurl}


\bibitem[\protect\citeauthoryear{Yeh, Zhu, Ulanova, Begum, Ding, Dau, Silva,
  Mueen, and Keogh}{Yeh et~al\mbox{.}}{2016}]%
        {Yeh.2016a}
\bibfield{author}{\bibinfo{person}{Chin-Chia~Michael Yeh}, \bibinfo{person}{Yan
  Zhu}, \bibinfo{person}{Liudmila Ulanova}, \bibinfo{person}{Nurjahan Begum},
  \bibinfo{person}{Yifei Ding}, \bibinfo{person}{Hoang~Anh Dau},
  \bibinfo{person}{Diego~Furtado Silva}, \bibinfo{person}{ABdullah Mueen},
  {and} \bibinfo{person}{Eamonn Keogh}.} \bibinfo{year}{2016}\natexlab{}.
\newblock \showarticletitle{{Matrix Profile I}: All Pairs Similarity Joins for
  Time Series: A Unifying View That Includes Motifs, Discords and Shapelets}.
  In \bibinfo{booktitle}{\emph{2016 IEEE 16th International Conference on Data
  Mining (ICDM)}}. \bibinfo{pages}{1317--1322}.
\newblock
\urldef\tempurl%
\url{https://doi.org/10.1109/ICDM.2016.0179}
\showDOI{\tempurl}


\bibitem[\protect\citeauthoryear{Yu, Jibin, and Jiang}{Yu
  et~al\mbox{.}}{2016}]%
        {Yu.2016}
\bibfield{author}{\bibinfo{person}{Qin Yu}, \bibinfo{person}{Lyu Jibin}, {and}
  \bibinfo{person}{Lirui Jiang}.} \bibinfo{year}{2016}\natexlab{}.
\newblock \showarticletitle{An Improved {ARIMA}-Based Traffic Anomaly Detection
  Algorithm for Wireless Sensor Networks}.
\newblock \bibinfo{journal}{\emph{International Journal of Distributed Sensor
  Networks}} \bibinfo{volume}{12}, \bibinfo{number}{1} (\bibinfo{date}{January}
  \bibinfo{year}{2016}).
\newblock
\urldef\tempurl%
\url{https://doi.org/10.1155/2016/9653230}
\showDOI{\tempurl}


\bibitem[\protect\citeauthoryear{Zhu, Imamura, Nikovski, and Keogh}{Zhu
  et~al\mbox{.}}{2017}]%
        {Zhu.2017}
\bibfield{author}{\bibinfo{person}{Yan Zhu}, \bibinfo{person}{Makoto Imamura},
  \bibinfo{person}{Daniel Nikovski}, {and} \bibinfo{person}{Eamonn Keogh}.}
  \bibinfo{year}{2017}\natexlab{}.
\newblock \showarticletitle{{Matrix Profile VII}: Time Series Chains: A New
  Primitive for Time Series Data Mining}. In \bibinfo{booktitle}{\emph{2017
  IEEE International Conference on Data Mining (ICDM)}}.
  \bibinfo{pages}{695--704}.
\newblock
\showISSN{2374-8486}
\urldef\tempurl%
\url{https://doi.org/10.1109/ICDM.2017.79}
\showDOI{\tempurl}


\end{thebibliography}

\end{document}